\documentclass[twocolumn]{aastex63}

\received{\today}
\accepted{AJ}

\shorttitle{Detection of Emission Line Galaxies in Filaments}
\shortauthors{Edwards et al.}

\graphicspath{{./}{figures/}}
\usepackage{subfigure}
\begin{document}

\title{Efficient detection of emission line galaxies in the Cl0016+1609 and  MACSJ1621.4+3810 supercluster filaments using SITELLE}

\correspondingauthor{Louise O. V. Edwards}
\email{ledwar04@calpoly.edu. 
\\Based on observations obtained at the Canada-France-Hawaii Telescope (CFHT) which is operated from the summit of Maunakea by the National Research Council of Canada, the Institut National des Sciences de l'Univers of the Centre National de la Recherche Scientifique of France, and the University of Hawaii. The observations at the Canada-France-Hawaii Telescope were performed with care and respect from the summit of Maunakea which is a significant cultural and historic site. The corresponding proposals were: 17BF001, 18BF001 and 19AF001. }

\author[0000-0002-9135-997X]{Louise O.V. Edwards}
   
\affiliation{California Polytechnic State University,
1 Grand Avenue, San Luis Obispo, CA, 93405, USA
}
\author[0000-0002-6991-4578]{Florence Durret}

\affiliation{Sorbonne Universit\'e, CNRS, UMR 7095, Institut d'Astrophysique de Paris, 98bis Bd Arago, F-75014 Paris}

\author[0000-0003-2629-1945]{Isabel M\'arquez} 
\affiliation{Instituto de Astrof\'isica de Andaluc\'ia, CSIC, Glorieta de la Astronom\'ia s/n, 18008, Granada, Spain}

\author{Kevin Zhang$^1$}

\begin{abstract}

We observe a system of filaments and clusters around Cl0016+1609 and
MACSJ1621.4+3810 using the SITELLE Fourier transform spectrograph at the Canada France Hawaii Telescope. For Cl0016+1609 ($z=0.546$), the observations span an $11.8\,$Mpc$ \times 4.3\,$Mpc region along an eastern filament which covers the main cluster core, as well as two $4.3\,$Mpc$\times4.3\,$Mpc regions which each cover southern subclumps. For MACSJ1621.4+3810 ($z=0.465$), $3.9\,$Mpc$\times 3.9\,$Mpc around the main cluster core is covered. We present the frequency and location of the emission line galaxies, their emission line images, and calculate the star formation rates, specific star formation rates and merger statistics. In Cl0016+1609, we find thirteen [OII]~3727\AA~ emitting galaxies with star formation rates between $0.2$ and $14.0\,$M$_{\odot}$ yr$^{-1}$. 91$^{+3}_{-10}$\% are found in regions with moderate local galaxy density, avoiding the dense cluster cores. These galaxies follow the main filament of the superstructure, and are mostly blue and disky, with several showing close companions and merging morphologies. In MACSJ1621.4+3810, we find ten emission line sources. All are blue (100$^{+0}_{-15}$\%), with 40$^{+16}_{-12}$\% classified as disky and 60$^{+12}_{-16}$\% as merging systems. Eight avoid the cluster core (80$^{+7}_{-17}$\%), but two (20$^{+17}_{-7}$\%)  are found near high density regions, including the brightest cluster galaxy (BCG). These observations push the spectroscopic study of galaxies in filaments beyond $z\sim0.3$ to $z\sim 0.5$. Their efficient confirmation is paramount to their usefulness as more galaxy surveys come online. 

\end{abstract}


\section{Introduction} \label{sec:intro}
Thirty years of observations \citep[e.g.,][]{Ramella92,Tempel14,Alvarez18},
dark matter (e.g., \citet{Springel08}), and cosmological simulations (e.g., \citet{Dolag06,Gheller16}) have revealed that mass in the universe is mostly confined to filaments with galaxy clusters located at their intersection.  These filaments and extensions are tracing the direction along which clusters are still accreting large amounts of baryonic gas \citep{Eckert15}, but the study of their galaxies lags far behind those in clusters because identifying filament systems requires very large fields of view which are populated mostly by interlopers.  \noindent

Recently, as statistical methods have been applied to large databases, over 15000 low redshift filaments ($z<0.2$) in the Sloan Digital Sky Survey (SDSS DR8) have been identified using these ``filament finders'',
showing $30-40\%$ of the galaxy luminosity is contained within these structures \citep{tem13}. At
more distant redshifts, fewer are known. Large extensions or filaments
have been found around twelve clusters of the DAFT/FADA
survey\footnote{http://cesam.lam.fr/DAFT/index.php} in the redshift
range $0.4<z<0.9$, based on density maps built from red-sequence
selected galaxies \citep{dur16}.  More filaments are sure to be found as next generation surveys from the Vera Rubin Telescope and results from the Dark Energy Spectroscopic Instrument (DESI) at the Mayall Telescope come online. However, to study the galaxy properties in detail is difficult as most of the star forming sources will be excluded from red-sequence maps, and as most sources in the field will be interlopers. An efficient way to locate and study the galaxies inside a filament is wide-field integral field spectroscopy.
\noindent

The driving objective of this paper is to trace star forming galaxies in large filaments at redshift $\sim0.5$, closer in time to the epoch of cluster collapse.  Some success at low redshifts ($z<0.3$) has been obtained using the starburst (SB) galaxies themselves as beacons. Strikingly, the highest specific star formation rates (sSFR), and the highest frequency of star forming galaxies in a superstructure are in the filaments \citep{Braglia07,Porter08,Edwards10,Biviano11,Darvish14,Bianconi16}. For local superclusters in the SDSS, \citet{coh17} finds that the frequency of star forming galaxies decreases with local galaxy density. Although, in line with the filament studies above, \citet{pog08} note it is specifically the moderate density environments (not the highest or lowest galaxy density regions) that have the highest frequency of line emitters. This is also in line with \citet{hom05}, \citet{mor05}, and \citet{ein18} who find emission line galaxies mostly located outside of the cluster virial radius. \citet{mah12} hypothesize that the increased activity is due to the moderate galaxy density in these regions, allowing for galaxy-galaxy interactions that lead to bursts of star formation and higher star formation rates \citep{par09}. The large-scale environment appears to matter for the intensity of star formation as well, as second only to galaxy mass, the GAMA survey has found the large-scale environment to modulate SFRs \citep{Alpaslan16}. Furthermore, \citet{vul10} and \citet{pau20} find lower emission line strengths for galaxies in high density environments, again at fixed stellar mass. But, the result is not consistent throughout the literature. \citet{pog08} find the SFRs and sSFRs are independent of local galaxy density, and \citet{ran20} find no trends in SFR with environment (cluster, field or filament, as described by \citet{dar17}).

The study of location, extent and morphology of star forming galaxies can test how galaxies might change as they interact with their environments. Ram pressure stripping as a galaxy falls into the cluster can cause a short burst before ultimately removing gas from the galaxy \citep{gun72}. Such cluster processes are expected to appear once the intracluster medium becomes significant. This idea can be tested by determining the location of starburst galaxies within well defined massive structures and by mapping starburst morphology. Harassment from gravitational encounters with other filament galaxies will also disturb the galaxies, and merger-driven bursts may use up the gas in a galaxy. Thus, searching for evidence of gravitational encounters - such as measuring galaxy asymmetry is also illuminating. 

This paper discusses line emitting galaxies in two superclusters from the DAFT/FADA survey, a project that, among other topics, traces clusters and their filaments in the $0.4<z<0.9$ redshift range by selecting red sequence galaxies \citep{Guennou10,dur16}. Cl0016+1609 is at a redshift of $z=0.546$, and  MACSJ1621.4+3810 is at $z=0.465$. 

\begin{figure*}[ht!]
\centering
\includegraphics[angle=0, width=0.95\textwidth, clip=true]{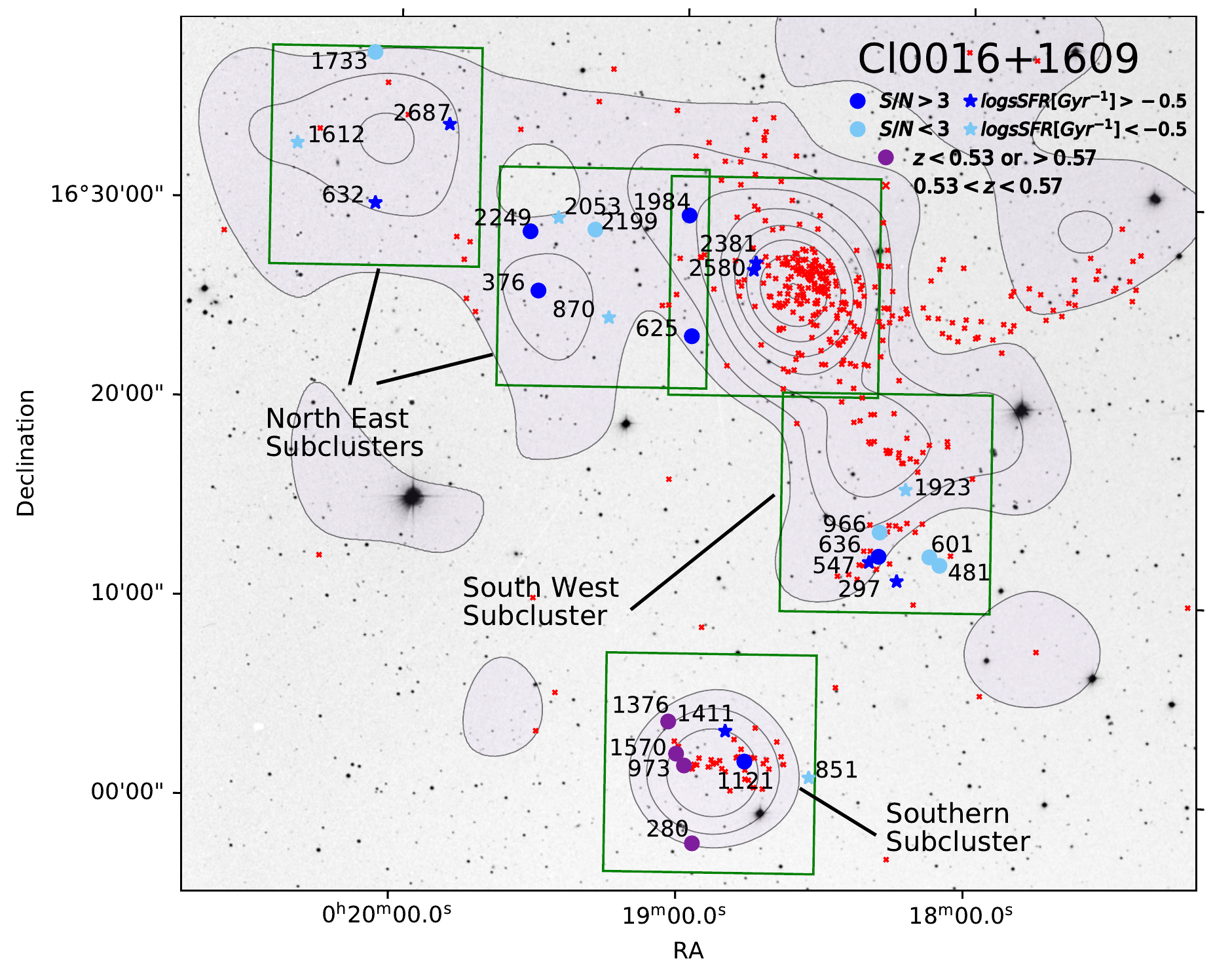}
\caption{Red sequence galaxy density map of the cluster Cl0016+1609
  and its surroundings. Grey contours show significance levels
  starting at $3\sigma$ above the background and increasing by $1\sigma$.  The five SITELLE
  fields observed are superimposed as green squares and measure $11^{\prime}\times 11^{\prime}$ ($3.1\,$Mpc$\times3.1\,$Mpc).  Small red crosses show the galaxies with a measured spectroscopic
  redshift in the approximate cluster redshift range ($0.53<z<0.57$). 
  The dark blue represents sources with emission lines that have $S/N>3$ and light blue sources have $S/N<3$. Circles represent galaxies with $log(sSFR) <-0.5\,$Gyr$^{-1}$, and stars represent galaxies with $log(sSFR) > -0.5 \,$Gyr$^{-1}$. 
  Purple circles show interloping emission line sources. North is up and East to the left.   
  }
 \label{CLzoneMap}
\end{figure*}

\begin{figure*}[ht!]
\centering
\includegraphics[angle=0, width=0.95\textwidth, clip=true]{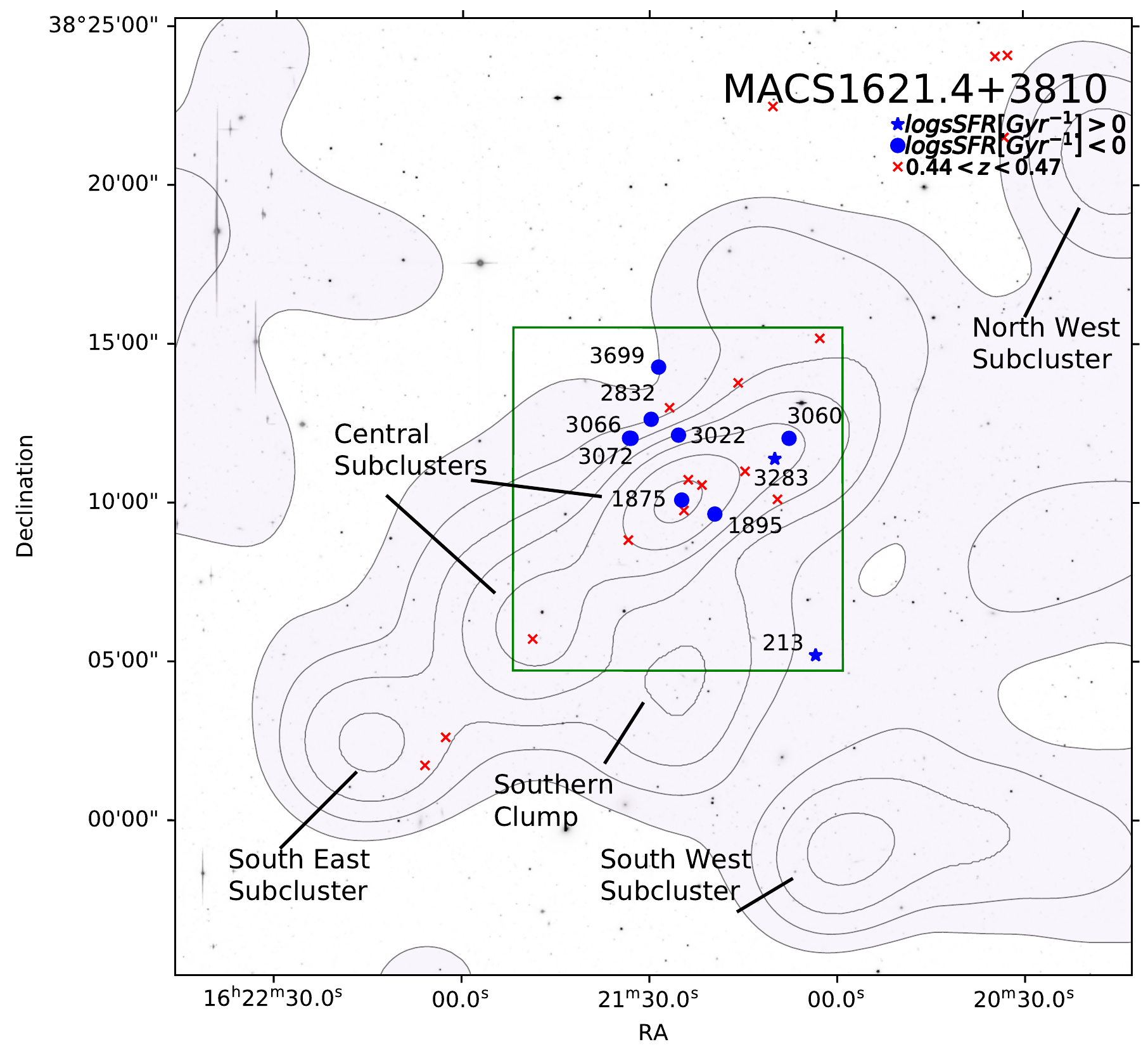}
\caption{Red sequence galaxy density map of the cluster MACSJ1621.4+3810 and its surroundings.  Grey contours show significance levels starting at $3\sigma$ above the background and increasing by $1\sigma$.  The SITELLE field is superimposed as a green square and measures $11^{\prime}\times 11^{\prime}$ ($3.9\,$Mpc$\times 3.9 \,$Mpc).  The small red crosses show the galaxies with a measured spectroscopic redshift in the approximate cluster redshift range ($0.44<z<0.47$).
The dark blue symbols indicate sources with emission lines. Circles represent galaxies with $log(sSFR) < 0.0 \,$Gyr$^{-1}$, and stars represent galaxies with $log(sSFR) > 0.0 \,$Gyr$^{-1}$. North is up and east is to the left. Note that sources with ID 3066 and 3072 are very close together and barely distinguished on the map.}
\label{MACSzoneMap}
\end{figure*}

Cl0016+1609 is one of the DAFT/FADA clusters with the most extensive filaments. Its galaxy density map reveals the presence of a main cluster core (Figure~\ref{CLzoneMap}) appearing embedded in a system of filaments over $7\,$Mpc long. Subclusters to the northeast and southwest, as well as a subcluster to the south are found. \citet{Higuchi15} measured virial masses for the main cluster core ($20.9 \pm 2.1) \times 10^{14}\,$M$_{\odot}$ and for the southern clump ($4.5\pm 1.1)\times 10 ^{14}$\,M$_{\odot}$. Several hundred redshifts have been gathered for this system (Figure~\ref{CLzoneMap}, red crosses). \citet{cra11} have measured the velocity dispersion of this cluster as $\sigma_{cl}=1490\,$km$\,$s$^{-1}$.

MACSJ1621.4+3810 is a less massive ($6.4\times 10^{14}\,$M$_{\odot}$) cool-core cluster \citep{edg03} which exists within a structure elongated in the NW-SE direction that extends over more than $7\,$Mpc. The SITELLE observations cover the two main central subclusters, and a portion of a southern clump. No data was obtained for the NW, SW and SE subclusters that can be seen in Figure~\ref{MACSzoneMap}. Less than two dozen redshifts are available for this system on NED\footnote{The NASA/IPAC Extragalactic Database (NED) is operated by the Jet Propulsion Laboratory, California Institute of Technology, under contract with the National Aeronautics and Space Administration.}.

By observing this large set of filament galaxies and those on the outskirts of subclusters, galaxies now entering the more dense environments are analyzed. Section~2 details the SITELLE observations. Section~3 presents the location of the emission line galaxies, testing the stripping hypothesis. In Section~4, the color, star formation rate (SFR) and morphology of the emission line galaxies are presented. Section~5 states the frequency of lime emitting galaxies. We conclude in Section~6.  Throughout the paper, physical sizes and distances assume H$_{o}=100h$~km~s$^{-1}$~Mpc$^{-1}$, $h=0.7$, $\Omega_{Vac}$, $\Omega_{M}=0.3$ and a flat universe.

\section{SITELLE OBSERVATIONS}

To find the star forming galaxies from the ground requires the detection of strong emission line sources. To characterize their extent and the  dynamics of the gas requires imaging the line emission over several angstroms, and to understand the shape of the galaxy and its surroundings requires deep imaging. These goals are concurrently realized with  integral field spectroscopy. The superclusters studied here are on the order of a degree on the sky. Thus, SITELLE, with its wide field of view provides a particularly efficient means of discovering the sources. 

SITELLE is an imaging Fourier transform spectrometer on the Canada France Hawaii Telescope \citep{gra12}. Its imaging field of view is $11^{\prime}\times 11^{\prime}$ with a spectral resolution from $R=2$ to $R>10^{4}$ which can be chosen by the user. The brightest cluster galaxy (BCG) of the local Perseus cluster \citep{gen18} and an examination of H$\alpha$ emission galaxies in Abell 2390 and Abell 2465, at $z=0.25$ \citet{Liu21} are currently the only other galaxy clusters to be observed by SITELLE. Thus, the objects identified in this paper represent the furthest systems thus far studied with the instrument.

Figure~\ref{CLzoneMap} illustrates the size and location of the 5 SITELLE pointings with respect to the location of the Cl0016+1609 structures. Data were taken over two observing runs. Proposal 17BF001 covers the cluster core and proposal 18BF001 covers two areas along the north-eastern filament, one along the southern filament and one on the southern subclump.

For MACSJ1621.4+3810, shown in Figure~\ref{MACSzoneMap}, there is one SITELLE pointing which covers the central $11^{\prime} \times11^{\prime}$ of the cluster, observed in semester 19A (19AF001).

\subsection{Instrument configuration} 
Both clusters were observed with the same configuration. In order to achieve a signal to noise of 5 in the emission lines, each field was observed for 4 hours\footnote{https://cfht.hawaii.edu/Instruments/Sitelle}. A resolution of 600 allows for a separation of the sky lines from the redshifted [OII]~3727\AA~ lines of interest. The sources were observed using $1\times1$ binning in order to avoid saturation, and as this was the most tested mode of observation for SITELLE. 


The data are flux calibrated with standard stars, wavelength calibrated with the HeNe lamp and phase calibrated using flat fields.

To capture the [OII]3726-3729 line emission, the C2 filter is used for Cl0016+1609 at $z=0.546$, and the C3 filter for MACSJ1621.4+3810 at $z=0.465$.

\subsection{Extracting emission line sources} 

Reconstructed deep images and detection maps are created with the SITELLE data using the ORCS package \citep{2015ASPC..495..327M}. Sources are then extracted using SExtractor \citep{Bertin96}. The deep image includes the sum of all emission in the filter, including the continuum emission, and so produces an image in which the galaxies can be easily identified. The detection map includes the addition of flux only from the emission lines within the filter's window, thus the morphology reveals the structure of the line-emitting gas, rather than the shape of the galaxy.  The output catalogs from both images are matched by RA and Dec to generate a list of emission line sources that have an optical galaxy host. For Cl0016+1609 there are 371 matched sources of the 2082 deep-detected galaxies with $r<21.5$, in the 5 detection frames. For MACSJ1621.4+3810 there are 304.

 \begin{figure}[!ht]
  \centering
\subfigure{\includegraphics[width=.9\linewidth]{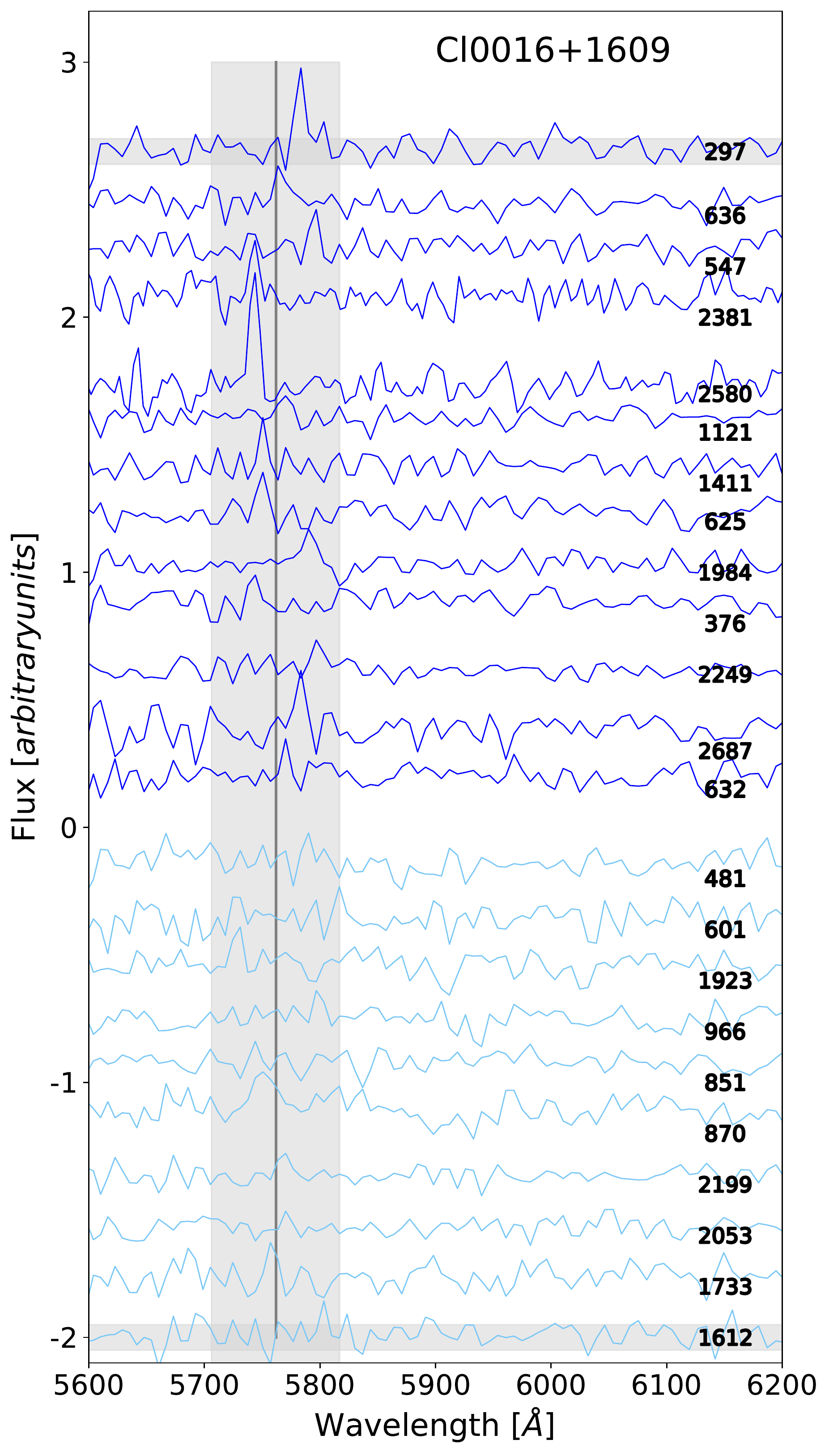}}  
  \label{CLspectra}
\subfigure{\includegraphics[width=.9\linewidth]{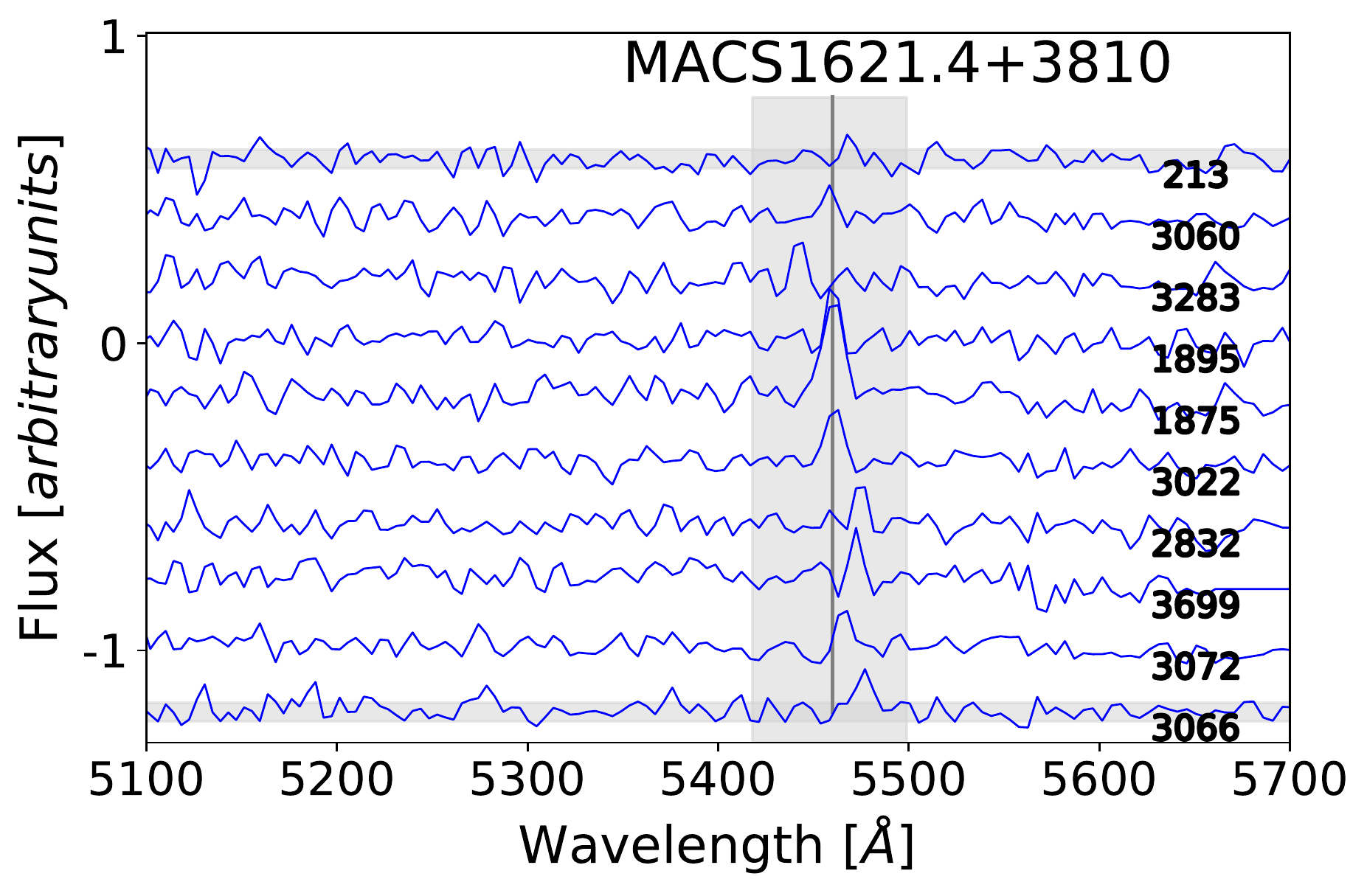}} 
  \label{MACSspectra}
\caption{The 23 emission line spectra for sources within the redshift range of Cl0016+1609 on the top, and the 10 for MACSJ1621.4+3810 on the bottom. Emission lines with $S/N>3$ are shown in dark blue and $S/N<3$ in light blue. The expected location of the [OII]3726-3729 doublet is shown with a vertical line, including a 3$\sigma_{cl}$ spread in grey. Flux levels are relative and error bars are the same for all spectra in a cluster, marked on the first and last galaxies.}
\label{clustspecs}
\end{figure}

For each source, a Python script subtracts strong sky lines by finding a nearby patch of sky, and a sky-subtracted spectrum for each object is generated. The generated object spectra are constructed by binning $3\times3$ pixels. This method includes the bright line emission from the target galaxies, and avoids any line emission from nearby sources, which are always more than 3 pixels from the source of interest. The resulting spectra are examined by eye by two of us (LE and KZ), and those sources with lines near the expected peak of the [OII]3726-3729 doublet are discussed below. 

For Cl0016+1609 there are 23 emission line sources within the expected redshift range of the supercluster, plus an additional four interloping systems identified. The source ID numbers and positions are listed in columns $1-3$, respectively, of Table~\ref{tab:objpropC} (top) and their spectra plotted in Figure~\ref{clustspecs}. Table~\ref{tab:objpropC} also shows the results of best-fit Gaussian functions used to model the lines and continuum noise. This includes the peak wavelength in column~4 and the standard deviation in column~5. The signal-to-noise level of the emission line is given in column~6. The noise is measured in the sky-subtracted continuum over in 50~\AA$\,$ bins that flank the location of the line peak. This error is illustrated in Figure~\ref{clustspecs} as a grey shaded region around the spectrum, for clarity, only the first and last spectra are marked. Where sample sizes are small, errors on percentages are derived from the posterior probability, assuming a binomial distribution and the Jeffreys interval {\it astropy} \citep{astropy}. For the 23 emission line systems within the expected location of the supercluster, 13 (56$\pm{10}$\%) have S/N greater than 3.0 and 10 (43$^{+10}_{-9}$\%) have low S/N.  There are four high S/N sources that have velocities which fall above or below 3$\sigma_{cl}$:  ID number 973 appears to be a foreground source identified in NED as SDSS J001858.83+160149.7, with a redshift of $z=0.1161$ \citet{tan07}. ID~1570 is identified as SDSSJ001900.37+160223.3 in NED, and is also a foreground source. There is no redshift included in NED, but using the H$\beta$--[OIII] complex that appears in the SITELLE data, its redshift is calculated to be $z=0.2137$. ID~1376, identified in NED as WISEA J001902.06+160350.0, is a background source. ID~280 has a strong emission line $>3\sigma_{cl}$, thus is included as a background object in this study.

The 10 identified emission line sources for MACSJ1621.4+3810 are shown in Table~\ref{tab:objpropC} (bottom) and plotted in Figure~\ref{clustspecs}. The identified emission line sources all have wavelengths within that expected for a cluster with a velocity dispersion $\sigma_{cl}<1100\,$km$\,$s$^{-1}$\citep{edg03}. All (100$^{+0}_{-15}$\%) have $S/N>3$. 

A NED search finds small levels of Galactic extinction for both sources, with $E(B-V)=0.05$ for Cl0016+1609, and $E(B-V)=0.01$ for MACSJ1621.4+3810. Similar to other SITELLE studies, we have applied no Galactic extinction correction \citep{dua19}. 
 
\begin{deluxetable*}{cccccccccccccc}
\tablenum{1}
\tablecaption{Emission line properties for Cl0016$+$1609  (top) and MACSJ1621.4$+$3810 (bottom).\label{tab:objpropC}}
\tablewidth{0pt}
\tablehead{
\colhead{ID} &  \colhead{RA} & \colhead{Dec} &\colhead{$\lambda_{obs}$} & \colhead{$\sigma_{obs}$} & \colhead{$S/N$} &
\colhead{SFR$\pm$err} &
 \colhead{color}&
  \colhead{log(M$_{*}/$M$_{\odot}$)}&
 \colhead{LGD}&
\colhead{G} &
 \colhead{M$_{20}$}&
 \colhead{Merg.}&
 \colhead{$R_{gal}/R_{em}$}\\
\colhead{} &  \colhead{[deg]} & \colhead{[deg]} &\colhead{[\AA]} & \colhead{[\AA]} & \colhead{} &
\colhead{[M$_{\odot}$ yr$^{-1}$]} &
 \colhead{}&
   \colhead{}&
 \colhead{Arb}&
\colhead{} &
 \colhead{}&
 \colhead{stat.} &
 \colhead{}
}\startdata
481 &4.524976 & 16.200258 & 5788.5 & 3.4 & 2.5 & 1.92 $\pm$  1.2 &1.51&10.08&2820 &$-$ &$-$ &$-$ &0.47\\
601 & 4.533818 & 16.207000 & 5817.0 & 3.0 & 2.2 & 1.85 $\pm$ 1.8 &2.06&10.23&2940 &$-$ &$-$ &$-$ &0.47\\
1923 &4.554944 & 16.263237 & 5728.0 & 1.8 & 2.8 & 1.74 $\pm$  0.2 &2.55&9.21&6610 &$-$ &$-$ &$-$ &0.76\\
297 &4.561975 & 16.187099 & 5782.8 & 4.3 & 5.8 & 2.29 $\pm$ 0.2 &1.21& 9.27&3430 &0.4898 &$-$1.6685 &$-$0.0740 &0.88\\
966 & 4.577526 & 16.227982 & 5799.9 & 1.7 & 2.5 & 8.92 $\pm$ 4.4 &2.34&10.67&4890 &$-$ &$-$ &$-$ &0.57\\
636 & 4.578539 & 16.208126 &5765.7 & 4.2 & 3.1 & 2.99$\pm$ 1.8 &1.21&10.17&4210 &0.4544 &$-$1.5991 &$-$0.0995 &0.70\\
547 & 4.585999 & 16.202838 &5793.8 & 1.7 & 4.2 & 8.93 $\pm$ 2.4 &0.92&10.34&4290 &0.4823 &$-$1.5564 &$-$0.0658 &0.38\\
851 & 4.640203 & 16.021096 & 5773.2 & 1.7 & 2.2 & 1.31 $\pm$ 0.3 &1.70&9.39&3800 &$-$ &$-$ &$-$&0.78\\
2381 &4.688636 & 16.451712 &  5743.5 & 4.2  & 4.8 &  2.47 $\pm$ 0.6 &1.31&9.77&19800 &0.4336 &$-$1.1836 &$-$0.0622 &0.88\\
2580 &4.690822 & 16.445283&  5743.1 & 4.3  & 7.2 & 14.00 $\pm$ 1.8 &1.23&10.28&20200 &0.4766 &$-$1.6657 &$-$0.0867 &1.00\\
1121 & 4.693259 & 16.035147 & 5768.3& 5.3 & 3.8 & 7.43$\pm$ 4.4 &1.47&10.65&8380 &0.4746 &$-$1.5519 &$-$0.0729 &0.90\\
1411 & 4.711034 & 16.060102 & 5750.9 & 3.5 & 5.0 & 1.28$\pm$ 0.2 &1.02&9.31&7300 &0.5513 &$-$1.6271 &$-$0.0073 &0.47\\
280$*$ & 4.738506 & 15.964865 & 5858.8 & 4.0 & 4.1 & 2.65$\pm$ 0.3 &$-$&$-$&4060 &$-$ &$-$ &$-$ &1.17\\
625 &4.743794 & 16.388556& 5749.8 & 4.5 & 3.8 & 7.29 $\pm$4.4 &2.60&10.53&4360 &0.5983 &$-$0.6288 &0.1779 &0.44\\ 
973$*$ & 4.745117 & 16.030534 & 5595.3 & 4.2 & 7.6 & 12.35 $\pm$ 1.6 &$-$&$-$&8600 &$-$ &$-$ &$-$ &0.50\\
1984 & 4.747106 & 16.489942&5790.8 & 6.1 & 3.9 & 3.27 $\pm$ 1.2 &1.00 &10.07&6940 &0.5325 &$-$0.9119 &0.0734 &1.00\\
1570$*$ & 4.751514 & 16.039738 & 6071.8 & 1.8 & 6.5 & 23.62 $\pm$ 11.3 &$-$&$-$&7650 &$-$ &$-$ &$-$ &1.17\\
1376$*$ & 4.758994 & 16.064248 & 6254.2 & 5.9 & 4.4 & 6.00 $\pm$ 0.8 &$-$&$-$&5320 &$-$ &$-$ &$-$ &0.50\\
870 &4.816326 & 16.402878& 5752.2 & 11.5 & 2.5 & 9.04 $\pm$  1.8 &3.49&10.18&4800 &$-$ &$-$ &$-$ &0.86\\
2199 &4.829372 & 16.477139& 5769.0 & 5.6 & 2.8 & 3.58 $\pm$ 2.4 &1.95&10.31&4580 &$-$ &$-$ &$-$ &0.44\\
2053 & 4.861224 & 16.485896 &5772.2 & 2.9 & 2.9 & 0.18 $\pm$ 0.2 &2.36&8.60&4110 &$-$ &$-$&$-$ &0.40\\
376 & 4.877579 & 16.424779&5743.2 & 6.2 & 3.1 & 13.85 $\pm$ 4.4 &1.17 &10.95&6110 &0.4154 &$-$1.5540 &$-$0.1317 &0.89\\
2249 & 4.886551 & 16.474815 &5798.3 & 4.5 & 3.3 & 1.82 $\pm$ 1.2 &1.02&10.00&4480 &0.3607 &$-$1.5795 &$-$0.1895 &0.73\\
2687 &4.958177 & 16.563301& 5783.0 & 4.7 & 4.7 & 2.10$\pm$  0.3 &1.31&9.40&5760 &0.5161 &$-$2.2380 &$-$0.1271 &1.60\\
632 &5.021137 & 16.496542& 5770.1 & 3.5 & 3.3 & 0.93 $\pm$ 0.2 &0.92&9.27&5640 &0.4621 &$-$1.6216 &$-$0.0950 &0.60\\ 
1733 &5.023732 & 16.622842& 5759.0 & 3.8 & 2.9 & 2.35 $\pm$ 1.2 &2.22&10.06&4090 &$-$ &$-$ &$-$ &0.47\\
1612 &5.090631 & 16.545895& 5802.9 & 3.4 & 2.7 & 0.78 $\pm$ 0.2 &1.48&9.12&5720 &$-$ &$-$ &$-$ &0.50\\
\hline
213  & 245.265365 &38.086847 &5469.7 & 1.4 & 3.3 & 0.32  $\pm$0.2 &$-$0.26&5.78&8090 &0.4947 &$-$1.2190 &$-$0.0066 &0.56\\
3060 & 245.286943&38.200770&5458.2 & 3.4 & 3.2 & 0.43  $\pm$0.2 &0.73&9.18&23200 &0.4325 &$-$1.3903 &$-$0.0920 &1.23\\
3283 & 245.291210 &38.190234&5442.2 & 1.2 & 3.6 & 0.87  $\pm$0.5 &0.38&8.35&23500 &0.5674 &$-$1.3865 &0.0420 &1.14\\
1895 & 245.331983&38.160703&5460.6 & 1.1 & 4.9 & 1.54  $\pm$0.2 &0.70&9.28&29900 &0.6404 &$-$1.5924 &0.0856 &0.86\\
1875 & 245.353178&38.169016&5460.2 & 6.1 & 6.2 & 22.79  $\pm$4.4  &1.36&11.56&44600 &0.5194 &$-$1.8768 &$-$0.0737 &1.67\\
3022 & 245.355679 &38.202744&5460.9 & 4.4 & 4.6 & 2.05  $\pm$1.2 &0.72&9.97&18300 &0.4521 &$-$1.5839 &$-$0.0996 &0.50\\
2832 & 245.373921 &38.210585&5474.7 & 1.2 & 3.6 & 2.68  $\pm$1.2 &1.06&9.97&9610 &0.6174 &$-$0.8767 &0.1624 &0.78\\
3699 & 245.377604&38.238522& 5472.4 & 3.2 & 4.8 & 0.69  $\pm$0.2 &0.25&9.12&7020 &0.4369 &$-$1.6092 &$-$0.1181 &0.78\\
3072 & 245.387344&38.200867& 5465.9 & 3.4 & 3.6 & 0.76  $\pm$0.4 &0.87&9.64&10800 &0.6350 &$-$1.3574 &0.1130 &0.78\\
3066 & 245.388434&38.201130&5476.6 & 4.3 & 4.2 & 0.62  $\pm$0.2 &0.66&9.03&10700 &0.6500 &$-$1.0269 &0.1737 &0.43\\
\enddata
\tablecomments{The object ID, RA and DEC are given in columns 1$-$3. The peak, standard deviation and signal to noise ratio are then listed, and they are measured on a best$-$fit Gaussian of the emission line. The mass-calibrated star formation rate and error is calculated using Equation~\ref{sfreqn2}. Column 8 lists the color information. Note that the four sources with asterisks fall outside the expected cluster redshift range, and no color information is calculated for these four. For Cl0016+1609, the color is ($g^{\prime}-i^{\prime}$) and for MACSJ1621.4$+$3810, the color is (V$-$I). Column 9 gives the stellar mass and column 10 gives the local galaxy density (LGD) in arbitrary units. Columns 11-13 list the shape information: The Gini coefficient (G), M$_{20}$, and the merging statistic, respectively. This information is only calculated for galaxies within the expected cluster redshift range that have a $S/N>3$. The final column lists $R_{gal}/R_{em}$, the measured radius of the galaxy continuum emission divided by the radius of the line emission.}
\end{deluxetable*}

\section{Location of Line Emitting Galaxies} 

In Cl0016+1609, the emission line galaxies follow the large scale structures and avoid the cluster cores. This is seen in Figure~\ref{CLzoneMap}.

For MACSJ1621.4+3810, the BCG is a line emitting galaxy. All other line emitting sources avoid the core.
 
To further quantify the environment of the emission line sources, the galaxy density at the location of the emission line source is identified as defined and measured in \citet{dur16}. Briefly, the cluster member galaxies are identified using the red sequence method \citep{Gla00} from deep ($r^{\prime}$ brighter than 24) color magnitude diagrams. The 2D density map is constructed by applying the adaptive kernel technique with a generalized Epanechnikov kernel \citep{sil86}. These are the same density maps shown as contours in Figures~\ref{CLzoneMap} and \ref{MACSzoneMap}.

\begin{figure}[!h]
\centering
\subfigure{\includegraphics[width=.95\linewidth]{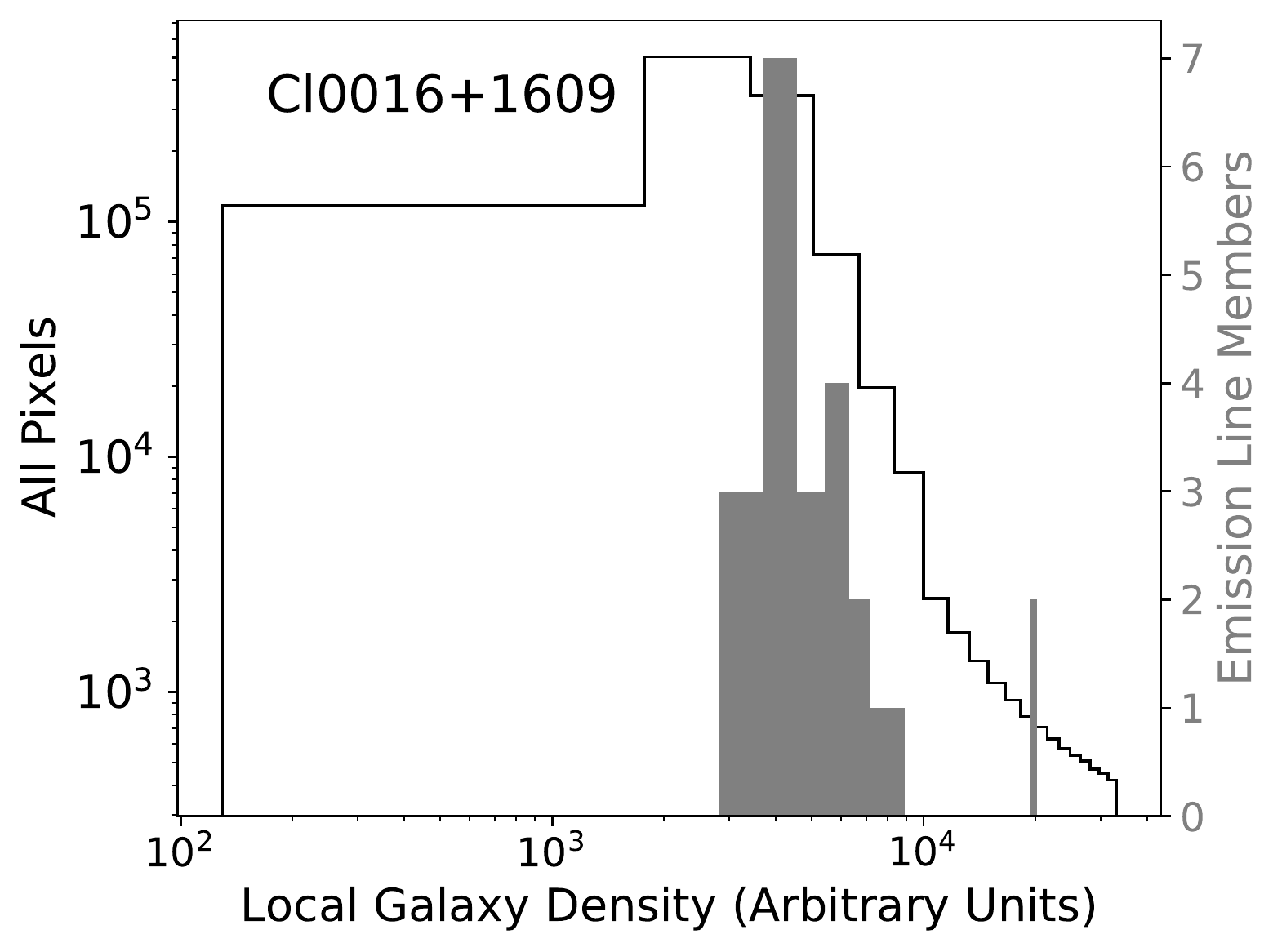}  }
\subfigure{\includegraphics[width=.95\linewidth]{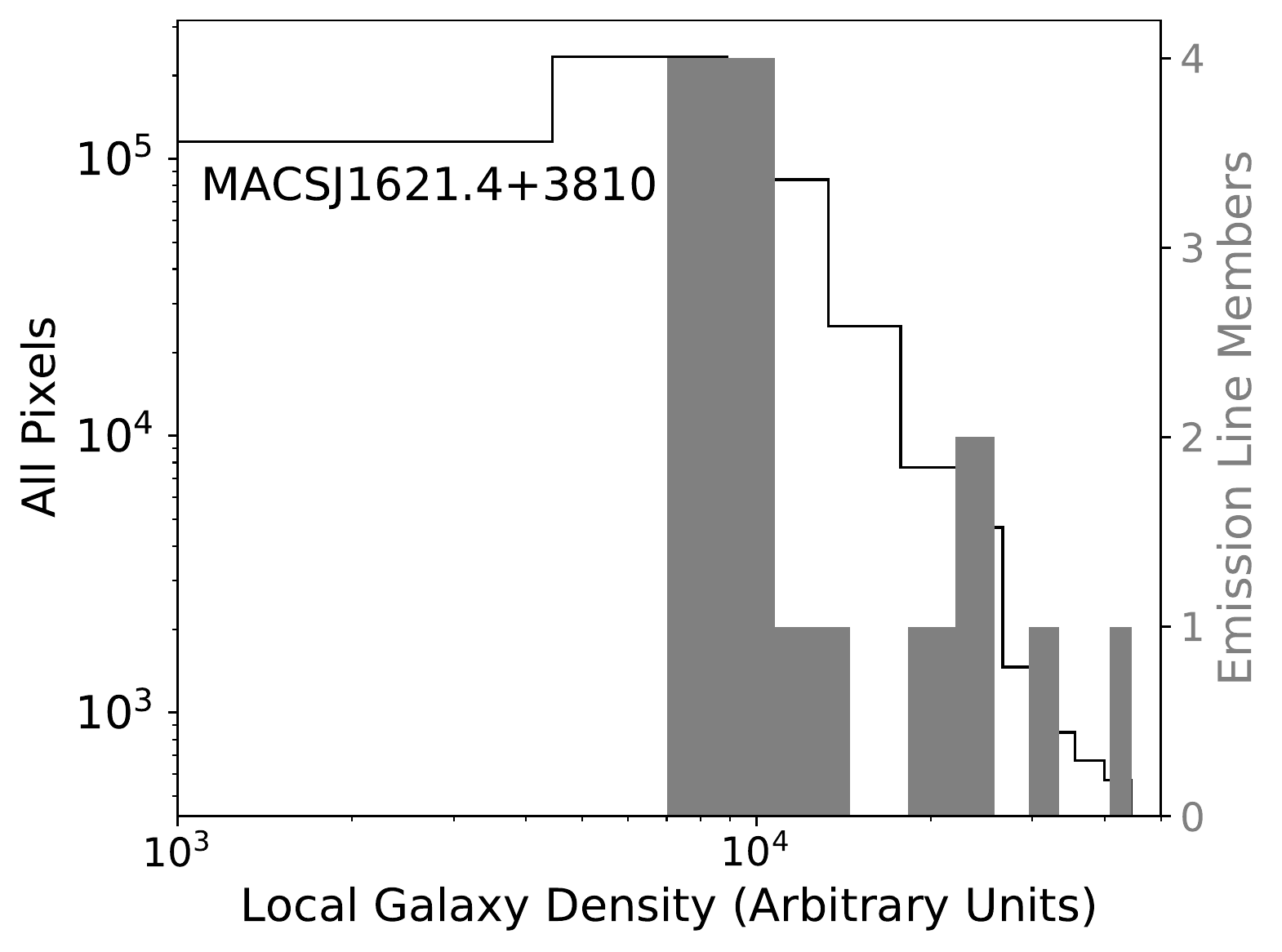}} 
\caption{Histogram of the local galaxy density (in arbitrary units) for the emission line objects (grey, filled) of Cl0016+1609 (top), and MACSJ1621.4+3810 (bottom). Alongside is a histogram of all pixels in the respective galaxy density image (black, unfilled). For Cl0016+1609, the star forming galaxies avoid the low density regions as well as the highest density regions of the cluster and subcluster cores. MACSJ1621.4+3810 has two within $LGD>25000$, one is the BCG. }\label{lgdhist}
\end{figure}

There is more empty space than dense structures, and for both Cl0016+1609 and MACSJ1621.4+3810, all emission line galaxies have a local galaxy density (LGD) value higher than the peak of galaxy density, as can be seen in the histogram of Figure~\ref{lgdhist}. The units of LGD are arbitrary and cannot be compared directly from one cluster to the other. But, by definition, these {\it moderate} density regions of the supercluster avoid both voids and dense cluster and subcluster cores. 

For Cl0016+1609, 21 of the 23 cluster emission line galaxies (91$^{+3}_{-10}$\%) are found in moderate galaxy density regions ($3000-8000$, arbitrary units) with two objects (9$^{+10}_{-3}$\%) found on the outskirts of the main cluster core (20000). This is consistent with the idea that galaxies go through an intense starburst as they enter the cluster and fall to the core \citep{bek99,bal00,fuj04} and is akin to what is seen at lower redshift \citep{mah11,pog04}.

Of the ten emission line sources for MACSJ1621.4+3810, eight (80$^{+7}_{-17}$\%) are found in moderate galaxy density regions ($7000-25000$), and two (20$^{+17}_{-7}$\%) are found near high density regions, including the BCG, at the peak of the galaxy density (44600). However, it is known that BCGs, especially in cool core clusters, often harbor bright emission lines \citep{cra99} and blue colors \citep{tru20} from star formation or AGN activity \citep[e.g.,][]{edw07,mcd10} likely related to their special place at the bottom of the potential, and their interaction with the intracluster medium \citep{fog17}.

For both superclusters, the different shapes of the emission line galaxy histograms in Figure~\ref{lgdhist} compared to all of the pixels in the 2D galaxy density maps suggests the set of line emitting galaxies is not simply a subset of the red cluster galaxies that are the source of the local galaxy density image. 

Most of the emission line galaxies in this sample are in the extended filaments, rather than cluster core, which supports the predictions of recent hydrodynamical simulations that point to filaments as effective sites of galaxy transformation \citep{lia19}.

\section{Properties of Emission line galaxies}

The emission line galaxies identified predominantly exist in the previously determined filament regions of the two superclusters observed. In this section, their colors, stellar masses, star formation rates and morphologies are measured and calculated. The data underscore the importance of merger-driven starbursts in moderate density environments. It is found that the emission line sources are much bluer than the red sequence galaxies of the clusters. Assuming the source of the line emission is star formation, moderate rates are calculated. These star forming galaxies are found to have a high frequency of nearby neighbors, spiral morphologies and show properties indicative of active merging. The emission line images often show whispy structures and material connected to nearby sources.

\subsection{Galaxy Color}

\begin{figure}[!t] 
\centering
\subfigure{\includegraphics[width=.99\linewidth]{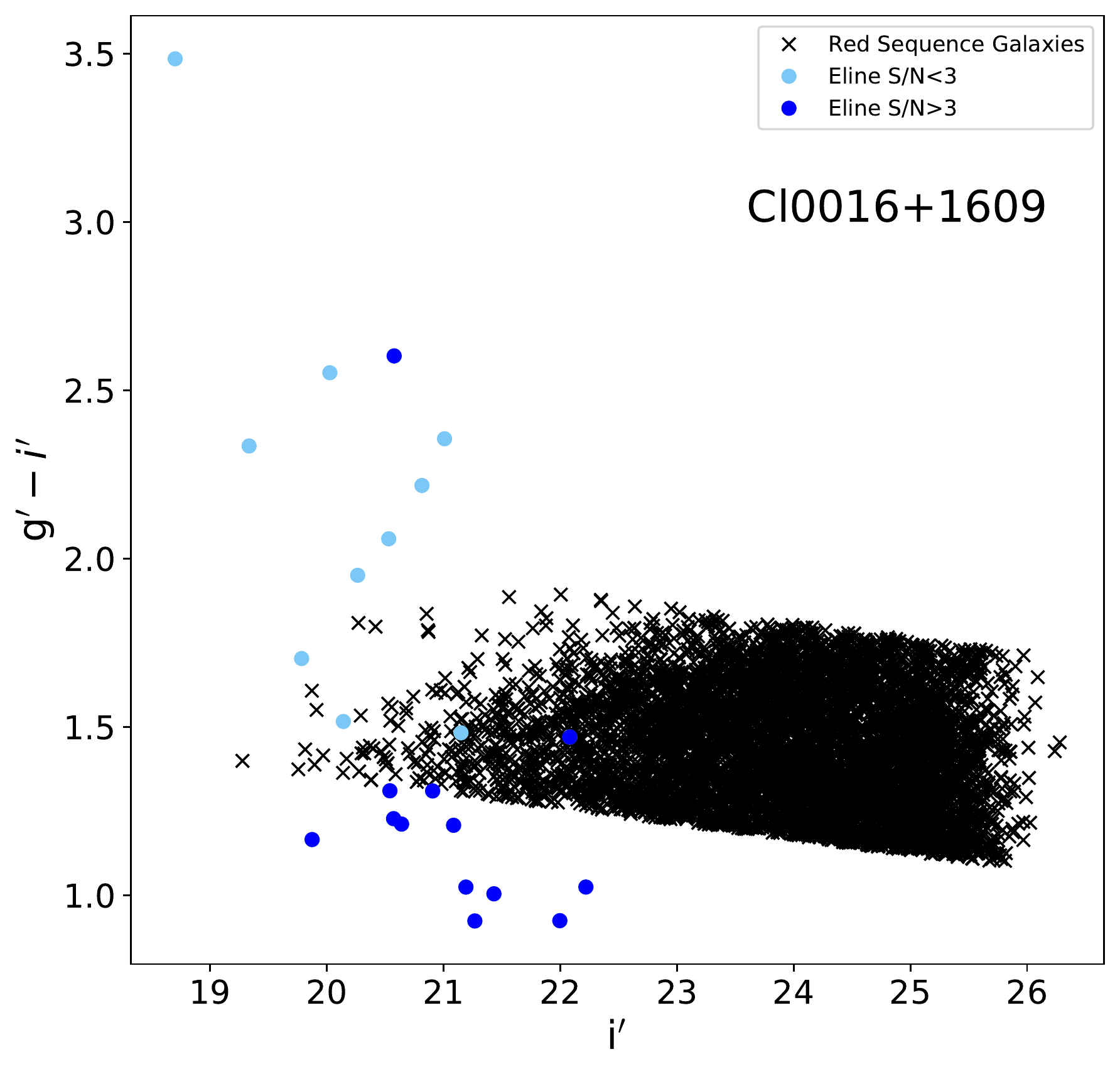} }
 \subfigure{\includegraphics[width=.99\linewidth]{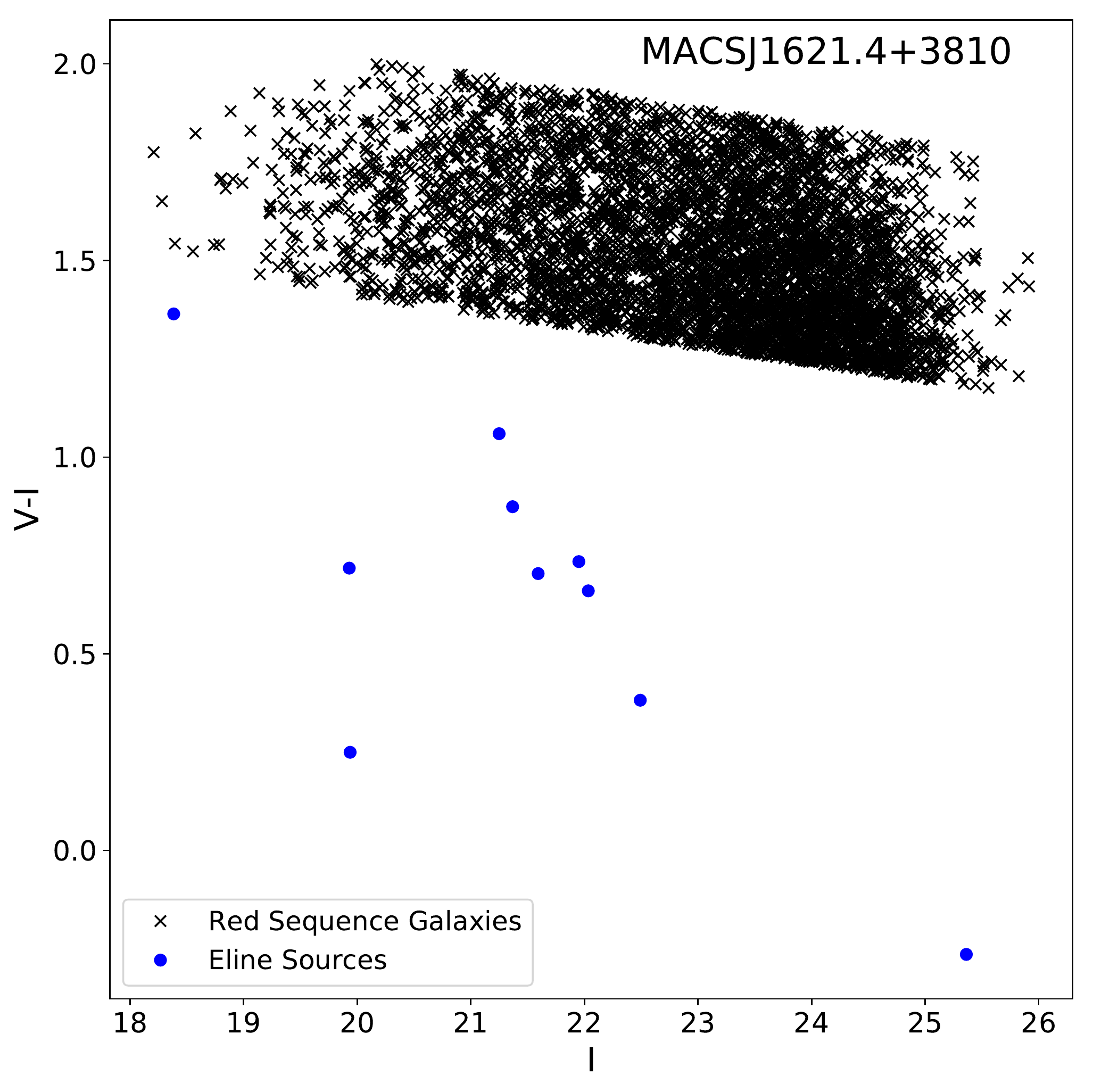} } 
  \caption{Color magnitude diagrams from \citet{dur16}. Top: The red sequence galaxies (x) of Cl0016+1609 and the emission line galaxies (low S/N as light blue points and high S/N sources as dark blue points). Of the 13 high S/N emission line galaxies, 11 are bluer than the red sequence.  Bottom: The red sequence galaxies (x) of MACSJ1621.4+3810 are plotted with the emission line galaxies (blue points). All emission line sources are blue. } \label{colmags}
\end{figure}

MegaCam images from the CFHT are used to determine the galaxy density maps for  Cl0016+1609, and Figure~\ref{colmags} (top), a color magnitude diagram for red sequence galaxies (crosses).  The emission line galaxies detected in this study (circles) are overplotted. All but two (85$^{+5}_{-15}$\%) of the high signal to noise emission line sources of Cl0016+1609 are blue galaxies, according to the red sequence definition of \citet{dur16}. The mean ($g^{\prime}-i^{\prime}$) color of the emission line sources is 0.4 magnitudes below the mean color of the red sequence galaxies. One source (ID$=$625) is very red with $g^{\prime}-i^{\prime}=2.6$, it has a very faint $g^{\prime}$ magnitude. This source may be a dusty red star forming galaxy \citep{fad08,gal09}, or an obscured AGN \citep{lac04,ima07}.

For MACSJ1621+3810, deep V and I imaging from the Subaru Telescope provide the color data. All matches are within 1$^{\prime\prime}$ except for ID=213, whose closest match is a I~$<25$ source at a distance of $4^{\prime\prime}$. For this system, the separation from the red sequence is even more extreme (Figure~\ref{colmags}, bottom), with 100$^{+0}_{-15}$\% of the emission line sources matched to blue galaxies. Even the BCG is blue, which is rare but not unique \citep{chu21}. The mean (V~$-$~I) colors of the emission lines are $\sim 1$ magnitude bluer than the red sequence. 

The emission line sources detected in this way show the expected blue color of strongly star forming galaxies \citep{bla03,bald04}, this gives confidence to the use of the Sitelle FTS  for identifying star forming galaxies in superclusters.

The k-corrected \citep{chi12} galaxy colors are used to calculate stellar masses, using the conversions in \citet{bel01} for the V and I colors and in \citet{bel03} for $g^{\prime}$ and $i^{\prime}$. They are listed in Table~\ref{tab:objpropC}. 

\subsection{Star formation rates}

Cosmic star formation rates have been plummeting since their peak around redshift 2 \citep{mad96,ly07}. By $z\sim0.5$, median star formation rates of $3.2 h^{-2}\,$M$_{\odot}$~yr$^{-1}$ are measured for blue galaxies, in cluster cores as well as the field \citep{coo08}.

Assuming the emission lines from these blue moderate density region sources are coming from star forming regions, the star formation rates can be calculated by calibrating the measured [OII]~3727\AA~ luminosity to an expected H$\alpha$ value, and applying the relation between star formation rate and H$\alpha$ luminosity \citet{ken98}. This requires knowing or assuming the extinction value at H$\alpha$, and the natural line ratio between the two. \citet{ken98} accomplished this by deriving a relationship for extinction-corrected [OII]~3727\AA~ luminosity, calibrated to line ratios of local blue-irregular, normal and peculiar star forming galaxies \citep{1989AJ.....97..700G,1992ApJ...388..310K}. Given the mass-metallicity relationship, higher metallicity galaxies are expected to have more dust, and thus more reddening. Concerned that high SFR systems would typically be underestimated, and low SFR systems overestimated, \citet{Kew04} derived a relationship for SFR based on [OII]~3727\AA~ luminosity that does not assume local reddening, so that one can apply a proper reddening correction. For cases where the reddening is unknown, the approach taken by \citet{gil10,gil11}, is useful. These authors use the galaxy stellar mass to empirically correct for reddening. The SFRs calculated in this paper use their equation (8), Equation~\ref{sfreqn2}, below.

\begin{equation} \label{sfreqn2}
SFR_{corr} = \frac{SFR_0}{a~tanh[(x-b)/c]+d}, 
\end{equation} 

\noindent where $x = log(M_{*}/M_{\odot})$ and $a = -1.424$, $b = 9.827$, $c = 0.572$, and $d = 1.700$. For sources with unknown mass, the calculations are performed assuming $log(M_{*}/M_{\odot})=10$. SFR$_{0}$ is the nominal star formation rate in M$_{\odot}$~yr$^{-1}$ before the mass-correction is applied, as defined in Equation~~\ref{sfreqn}, below.

\begin{equation} \label{sfreqn}
SFR_0 = \frac{10^{0.4 A_{H\alpha}}}{1.5 r_{lines}} \frac{L([OII])}{1.27 \times 10^{41}}, 
\end{equation} 

\noindent where L([OII]) is the [OII]~3727\AA~ luminosity in erg~s$^{-1}$, r$_{lines}$ is the ratio between the [OII]~3727\AA~ and H$\alpha$ emission lines, taken to be  0.5, and A$_{H\alpha}$ is the extinction at H$\alpha$. A value of A$_{H\alpha}$=1.0$\,$mag is adopted to calculate the nominal rate, SFR$_0$, and then the correction is applied (following \citet{gil10}). 

There are several important caveats worth mentioning here. First, AGN are not removed. If there is AGN contamination, the values quoted here would be overestimates \citep{zhu19}. However, AGN are thought to be most important if found along the red sequence \citep{yan06}, whereas most of the sources in this study avoid the red sequence. Second, the calibration of the relationship is based on galaxies between 0.03$<$z$<$0.20. Several studies have found that SFRs increase with redshift at fixed stellar mass. For example, total SFRs calculated using UV and FIR data increase with redshift at fixed stellar mass \citep{tom16}, so mass-calibrated SFRs may still be underestimated by a factor of $\sim$5 given the higher redshift of the clusters studied in this paper. Third, there is a huge variation source-to-source in the extinction levels at a given stellar mass. \citet{mou06} calculate the scatter in SFR is a factor of 2.5. Column 7 of Table~\ref{tab:objpropC} gives the calculated star formation rate and its error. The error quoted is calculated from the 1$\sigma$ uncertainties in the [OII]~3727\AA-derived SFR at each mass, as compared to the corrected H$\alpha$-derived rates (see Figure~3 of \citet{gil10}).

\begin{figure}[!h]
\centering
\subfigure{\includegraphics[width=.99\linewidth]{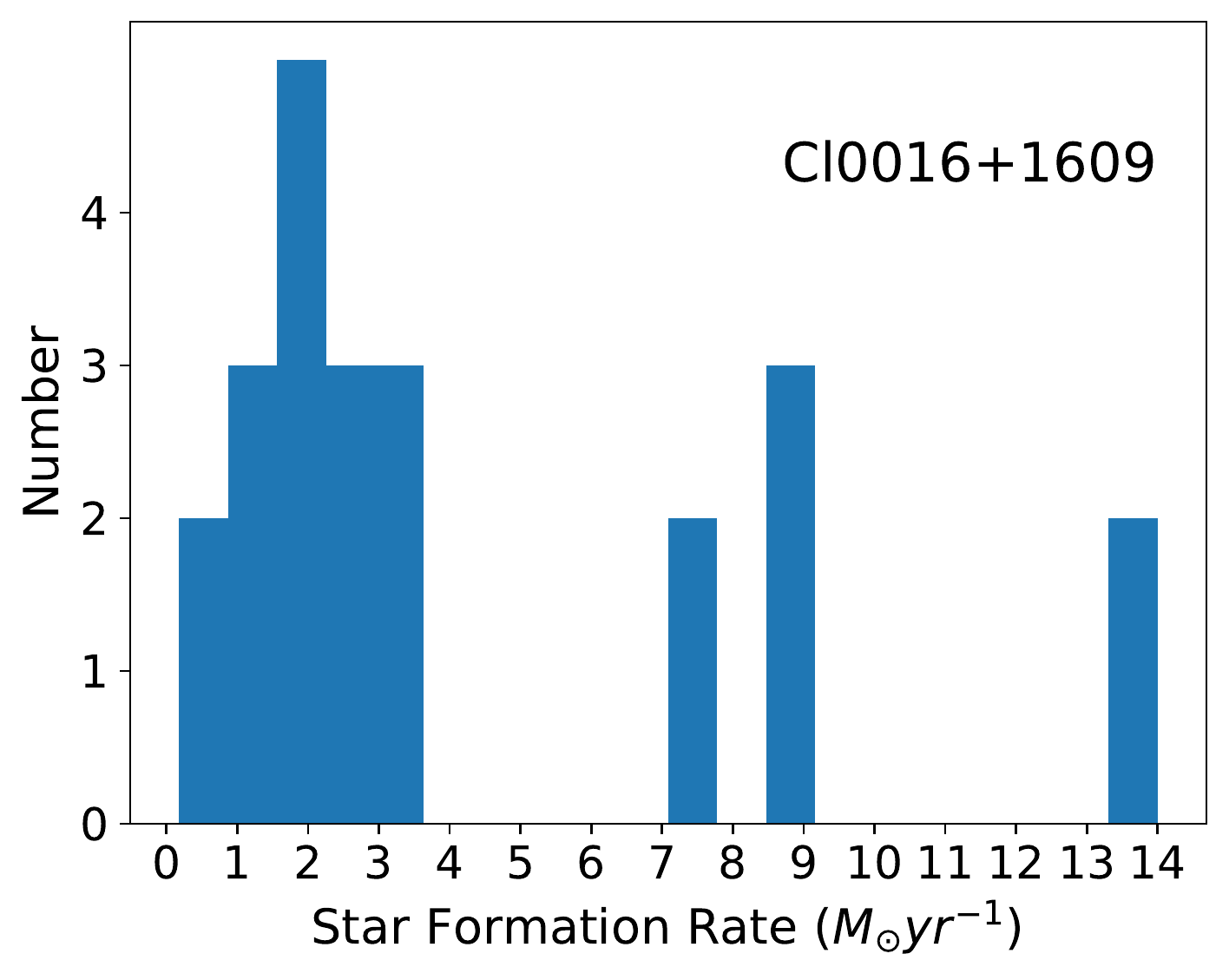}  }
\subfigure{\includegraphics[width=.99\linewidth]{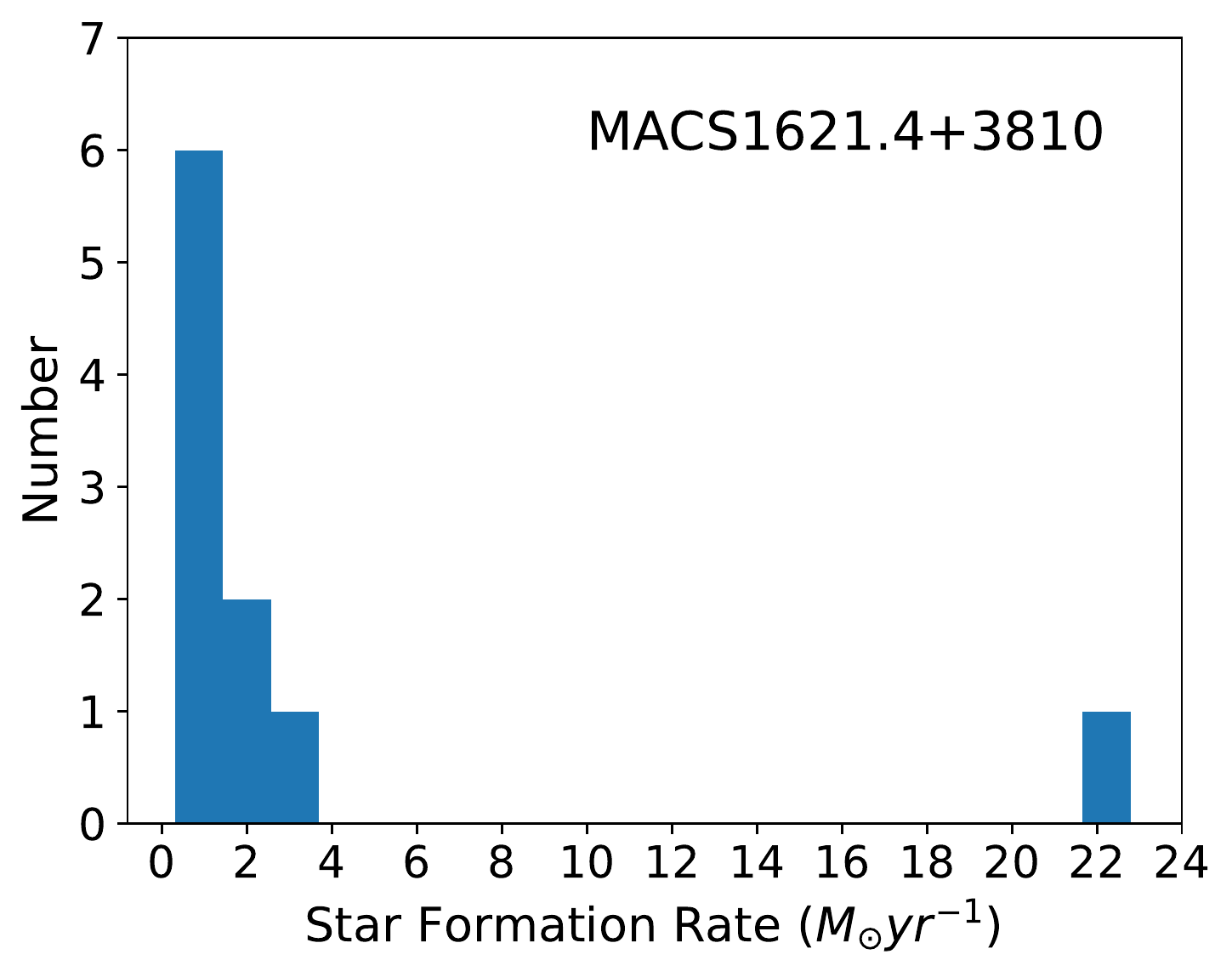} } 
\caption{Histogram of the star formation rates} for the emission line objects. Cl0016+1609 is shown on the top and MACSJ1621.4+3810 is shown on the bottom. The most common star formation rate is $\sim2 \,$M$_{\odot}$~yr$^{-1}$.\label{sfrhist}
\end{figure}

For Cl0016+1609, the star formation rates range from $0.2$ to $14\,$M$_{\odot}$~yr$^{-1}$, with a median star formation rate of $2\,$M$_{\odot}$~yr$^{-1}$ as can be seen in the histogram of Figure~\ref{sfrhist}. The star forming galaxies are blue when compared to the cluster members (Figure~\ref{colmags}), with one exception (ID$=$625), which is perhaps an AGN or an extremely dusty starburst.

\begin{figure*}[!ht]
\centering
\subfigure{\includegraphics[width=.5\linewidth]{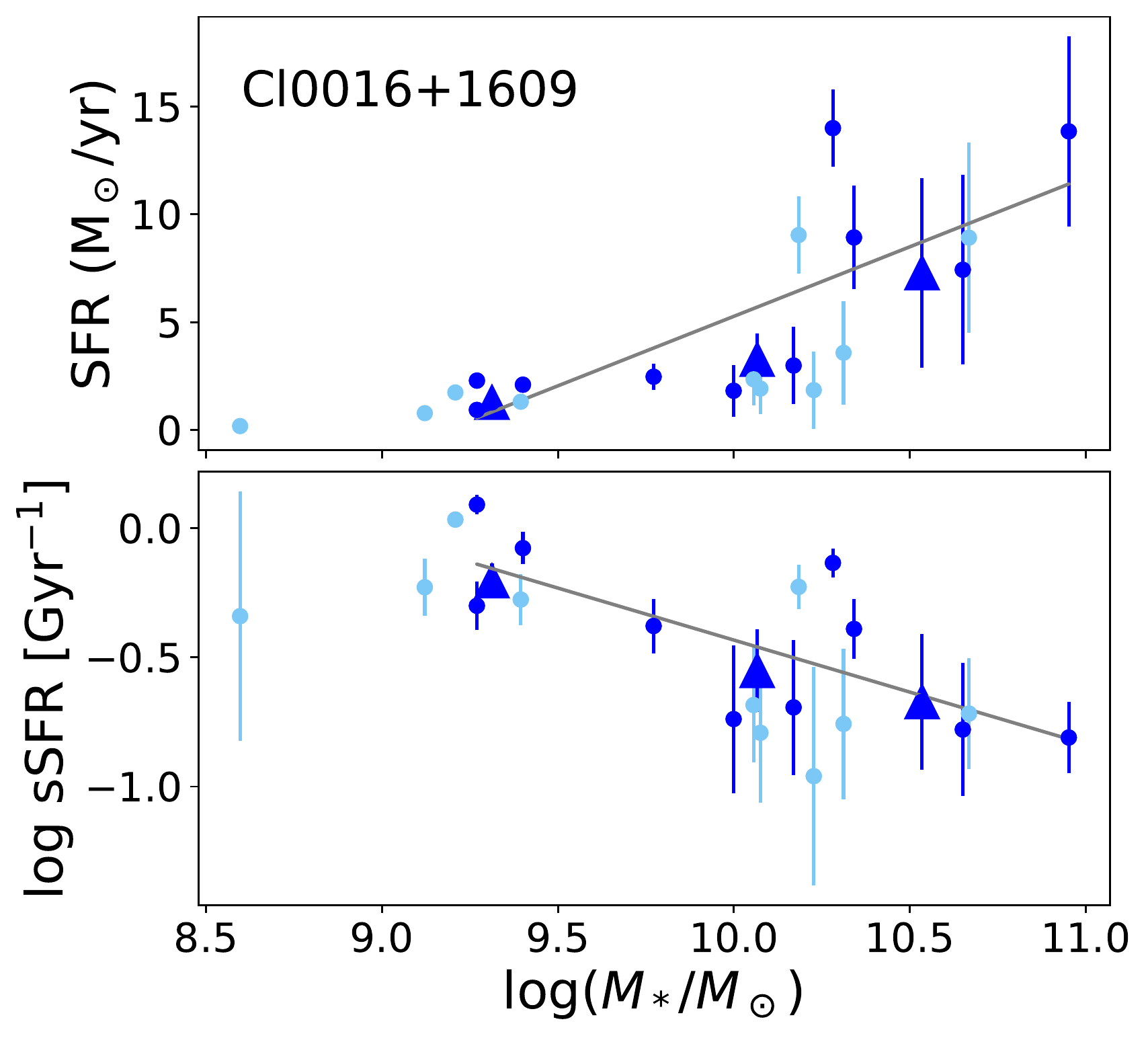} } 
\subfigure{\includegraphics[width=.47\linewidth]{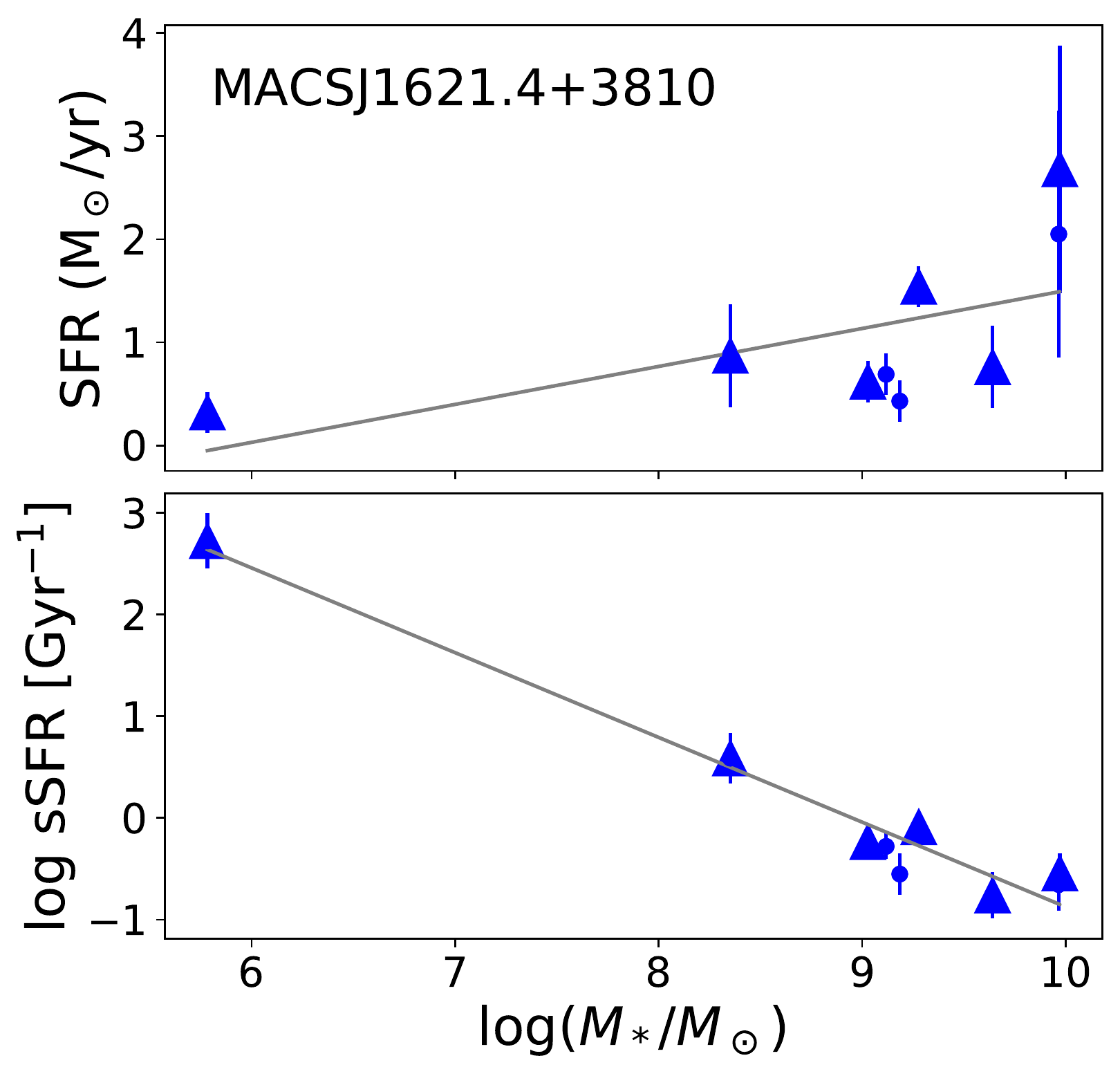}  }
\caption{Top: Relationship between the star formation rate (top) and specific star formation rate (bottom)} with $log(M_{*}/M_{\odot})$. Cl0016+1609 is shown on the left, and MACSJ1621+3810 on the right. Triangles label the merging systems (see text for details). \label{rprops_sfr}\label{rprops_sfrcl}
\end{figure*}

The two highest star formation rate galaxies have peak wavelengths close to the expected location for the cluster, bright $i^{\prime}$ magnitudes, small isophotal areas and radii, and thus bright $i^{\prime}$ surface brightnesses (see Table~\ref{tab:objpropC}). Their luminosity is emitted from a concentrated region. No evidence of a relationship with the galaxy position angle, ellipticity or elongation is identified. The set of star forming galaxies as a whole shows no strong trend with many of the galaxy properties measured (local galaxy density, velocity difference between the galaxy and cluster, surface brightness and radius). But, the fainter $i^{\prime}$ magnitude galaxies and those with bluer ($g^{\prime}-i^{\prime}$) colors have lower SFRs. The SFRs and specific star formation rates (sSFR) show the expected trend, with higher star formation rates for high mass galaxies, and higher specific star formation rates for low mass galaxies (see Figure~\ref{rprops_sfrcl}, left panels. The R$^{2}$ statistic for the fits is R$^{2}=0.62$ and R$^{2}=0.58$, respectively). The sSFR appears evenly distributed across values of LGD.

For MACSJ1621.4+3810, the star formation rates range from 0.32 to $22.8\,$M$_{\odot}$~yr$^{-1}$, with a median value of $0.8\,$M$_{\odot}$~yr$^{-1}$ (Figure~\ref{sfrhist}, bottom). In this cluster, the source with the strongest emission lines, and highest calculated star formation rate is the BCG  (ID$=$1875). This source is the reddest of the emission line sources, though still 0.3 magnitudes bluer than the red sequence in (V~$-$~I) (see Figures~\ref{colmags} and~\ref{rprops_sfr}). BCG's emission lines are usually associated with the cluster's cool core, so this source is removed when looking at possible trends with SFR. With only 9 data points for the emission line sources left, it is difficult to assert strong trends in the data. Nonetheless, a least squares fit to LGD and mass with SFR and sSFR was calculated. There is no trend with local galaxy density, but a weak trend with SFR increasing with mass is found (R$^{2}=0.34$), and the sSFR steadily decreases with mass (R$^{2}=0.96$; see Figure~\ref{rprops_sfr}, right).   The results are in line with the strong correlation found between SFR and galaxy stellar mass for local galaxies \citep{bri04, lar10, rod08, vul19}. 

At lower redshift, Coma, Abell~1763 and Abell~901 \citep{Edwards10,Bianconi16,gal09} also host the highest sSFRs in the filaments. \citet{mor05} find a small set of [OII]~3727\AA~ emitting galaxies surrounding the main core of Cl0024+16 at $z=0.4$. Those galaxies host a spread of [OII]~3727\AA~ luminosity, and, as in these superclusters, no trend with LGD is measured.

\subsection{Morphology of Emission line galaxies}

If the star formation activity of galaxies residing in filaments is indeed a result of galaxy-galaxy interactions, they might be expected to show tell-tail  signs of merging like nearby neighbors, tidal tails or trainwreck shapes. 

Indeed, deep images from the SITELLE data of all the emission line sources are constructed (left panel of Figures~\ref{cl_asym1} and  \ref{cl_asym2} for Cl0016+1609 and left panel of Figures~\ref{macs_asym1} and \ref{macs_asym2} for MACSJ1621.4+3810). Eight of the 23 emission line galaxies for Cl0016+1609 (35$^{+11}_{-9}$\%) show nearby projected companions within $50\,$kpc (IDs:  2381, 2249, 1984, 2687, 297, 636, 1411, 1211). For MACSJ1621.4+3810, half of the 10 emission line systems (50$\pm{14}$\%) host nearby projected companions (IDs: 213, 1875, 3022, 3066 and 3072). A few also display asymmetric shapes (Cl0016+1609: 2381, 376; MACSJ1621.4+3810: 213, 3022, 3699), emission-line images revealing multiple cores (MACSJ1621.4+3810:3022) and material connecting to nearby systems (e.g. MACSJ1621.4+3810:213, 1875).

In our two superclusters, 39$^{+9}_{-8}$\%  of emission line sources have nearby neighbors, this fraction is similar to that for luminous infrared galaxies (LIRGs)  \citep{shi09}, and significantly greater than $4-20$\% usually found at $z\sim 0.5$ \citep{lot11} either by close-pair analysis \citep{lin08}, the Gini coefficient \citep[explored below]{lot08}, or Asymmetry \citep{shi09}. The high fraction of nearby neighbors is also in line with the studies of \citet{guo2015} who find that galaxies in filaments have up to a factor of 2 more satellites than those outside of filaments. 

\begin{figure*}[!ht]
\centering
\subfigure{\includegraphics[width=.99\linewidth]{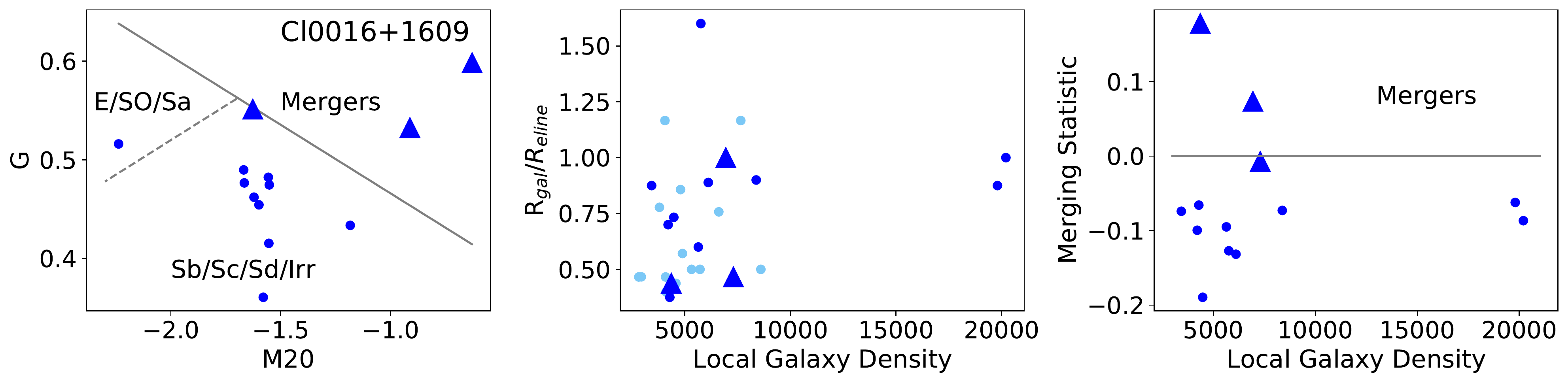} } 
\subfigure{\includegraphics[width=.99\linewidth]{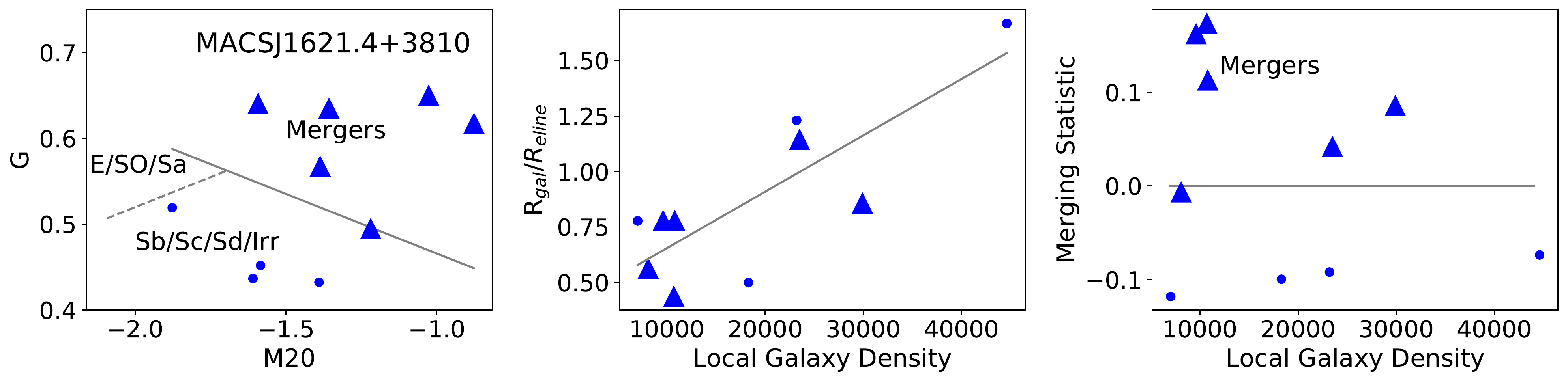}  }
\caption{Properties of the emission line galaxies and their environments. Cl0016+1609 is shown on the top, and MACSJ1621+3810 is shown on the bottom. Left panels: The G - M$_{20}$ values separate merging systems from non-merging systems. Middle: the ratio of the galaxy radius to the emission line region radius is plotted as a function of local galaxy density. Right: the merger statistic is plotted against the local galaxy density. Merging systems are found in both clusters, and reside in moderate density environments. Light blue colors represent $S/N<3$, dark blue represents $S/N>3$, and merging systems are shown as triangles. }\label{cl_rprops}
\end{figure*}

\begin{figure*}
\centering
\mbox{\subfigure{\includegraphics[width=3.6in]{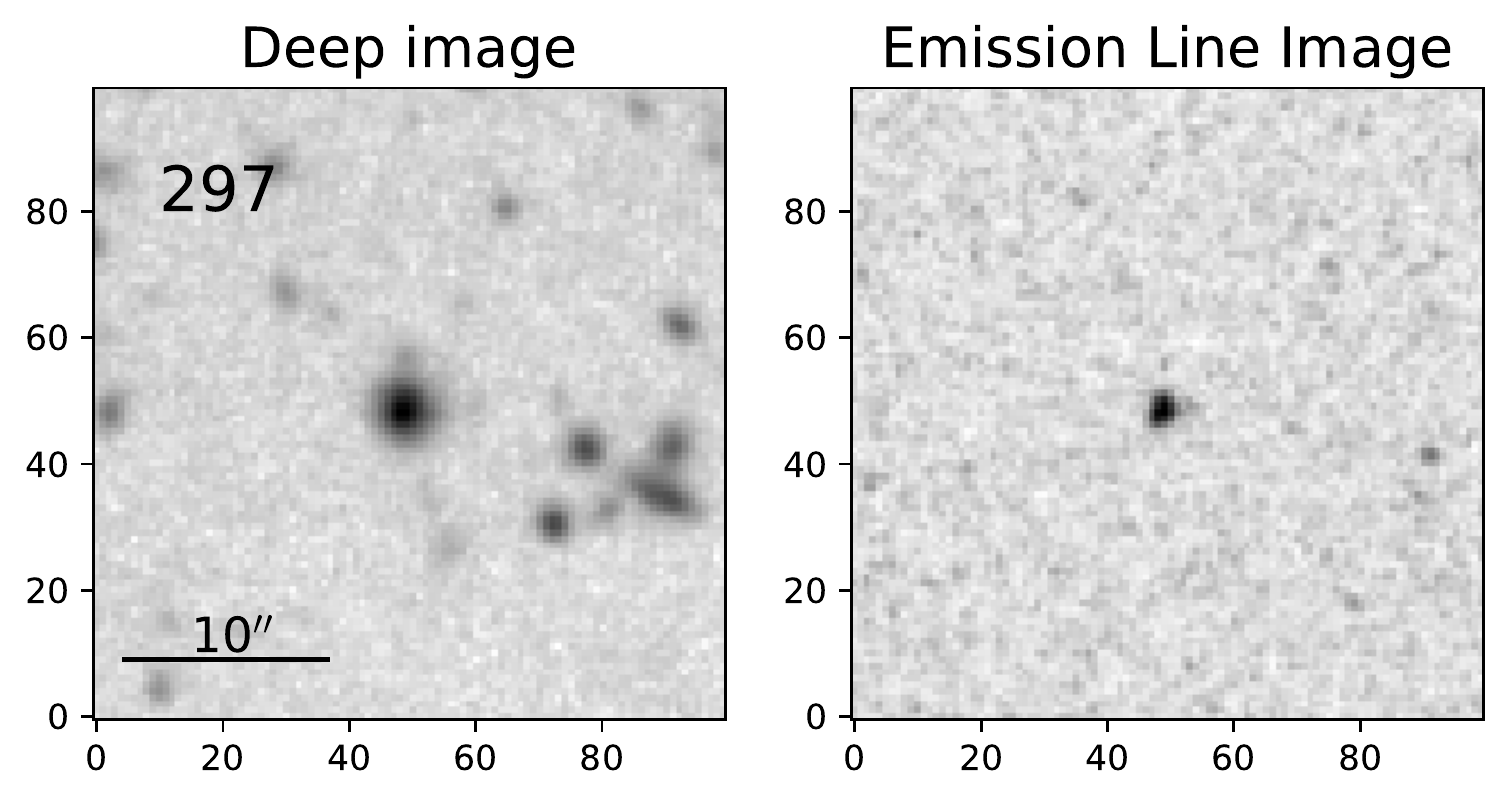}}\hfill
\subfigure{\includegraphics[width=3.6in]{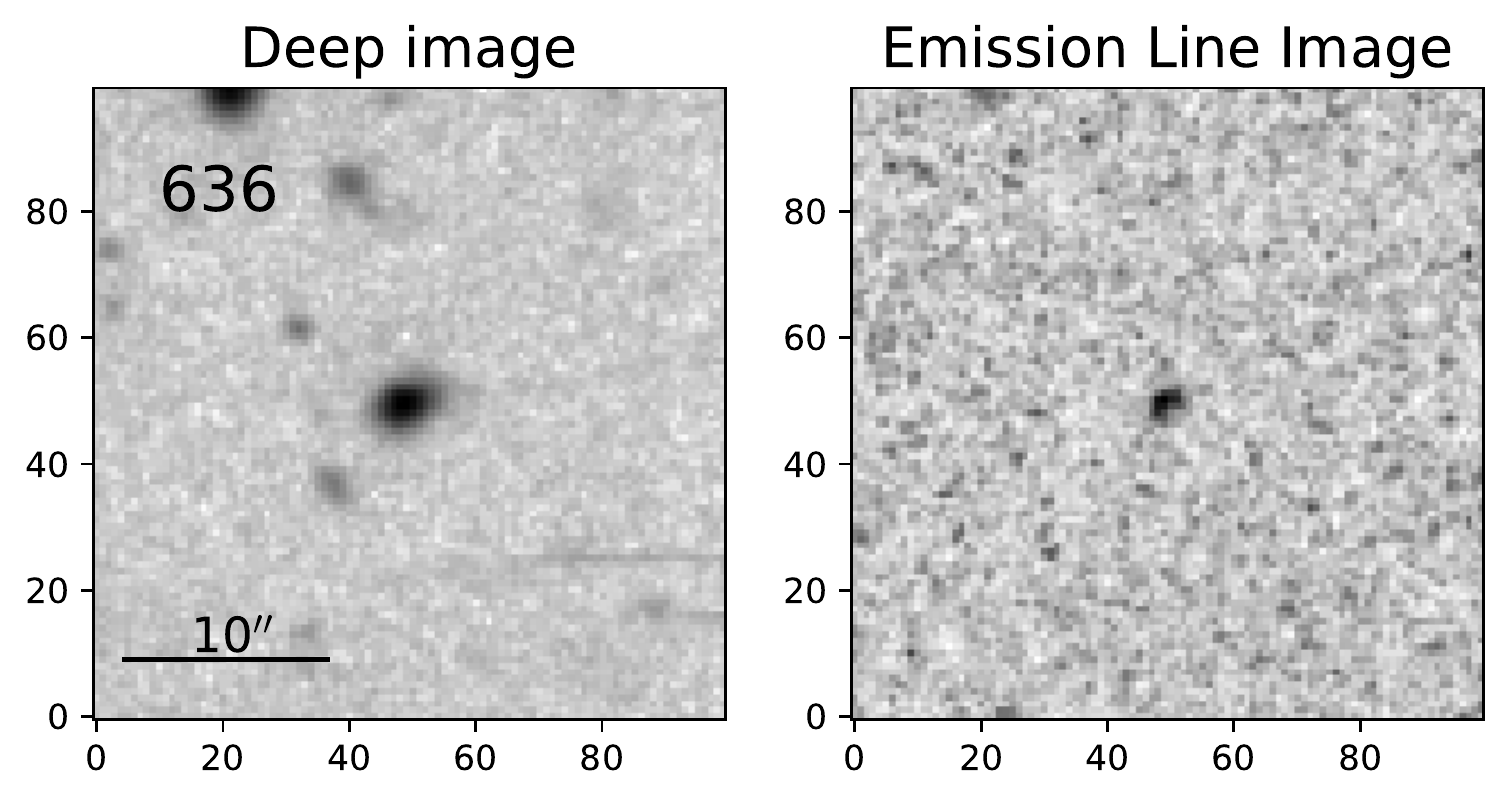}}}
\mbox{\subfigure{\includegraphics[width=3.6in]{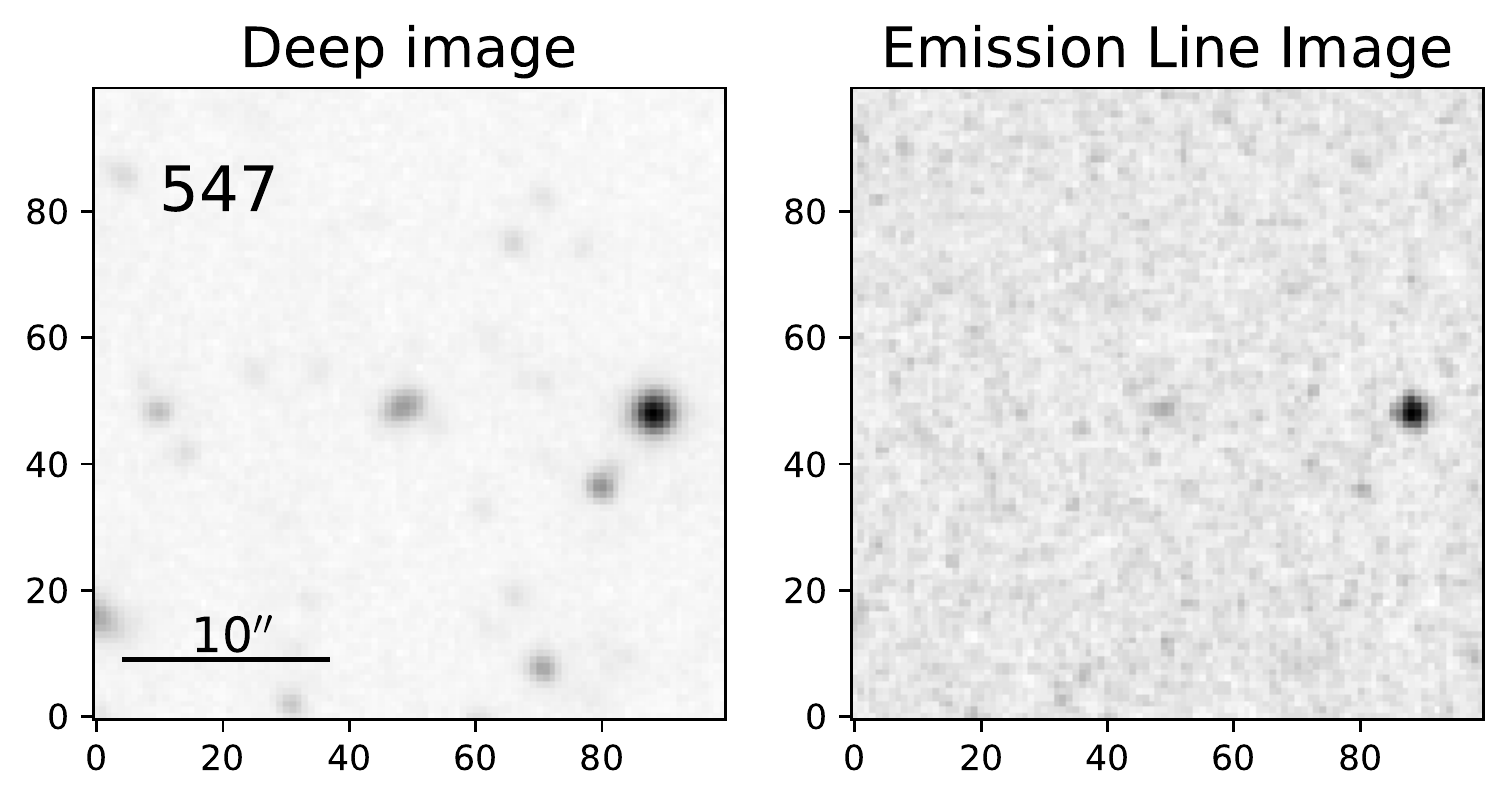}}\hfill
\subfigure{\includegraphics[width=3.6in]{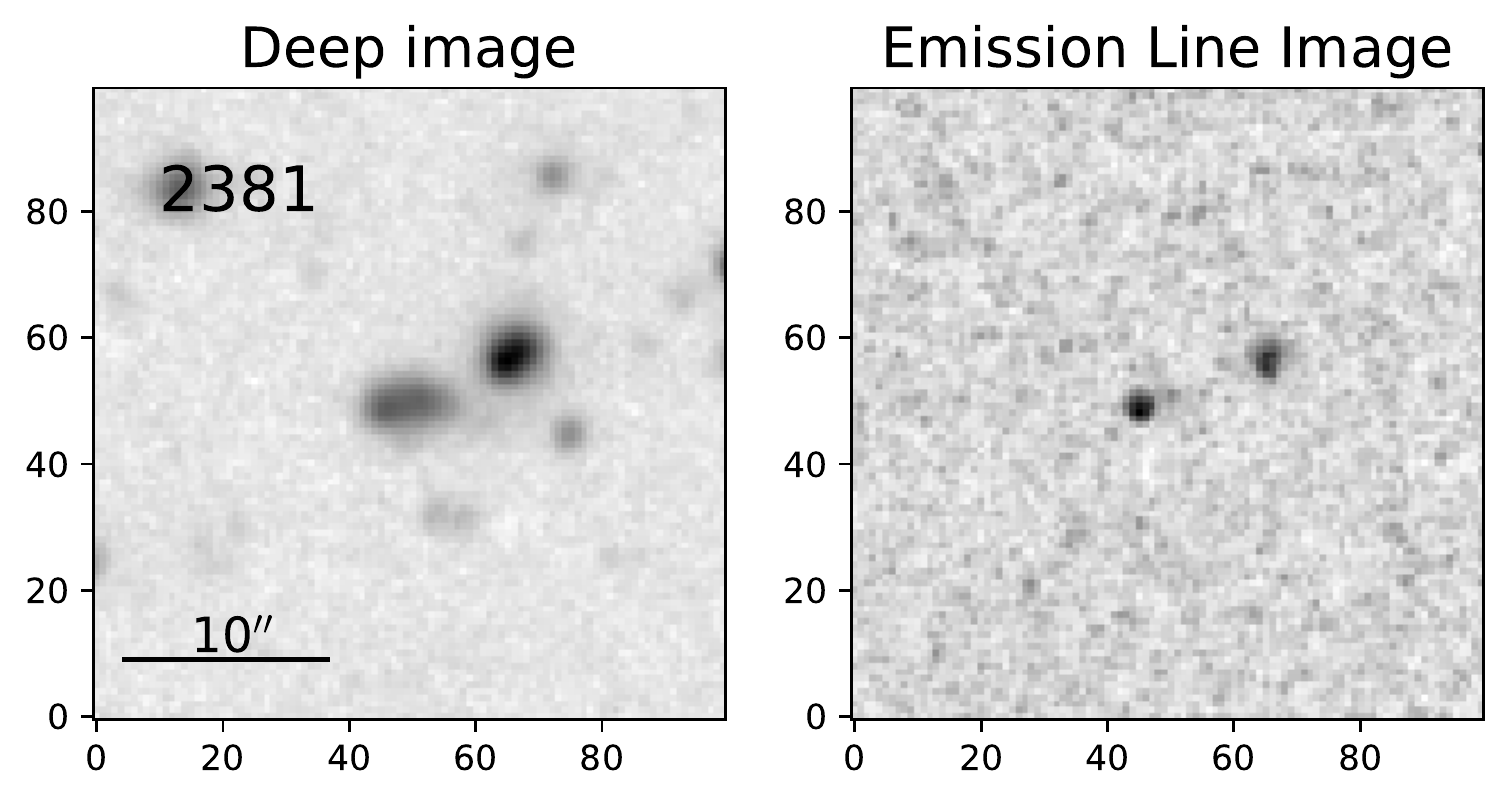}}}
\mbox{\subfigure{\includegraphics[width=3.6in]{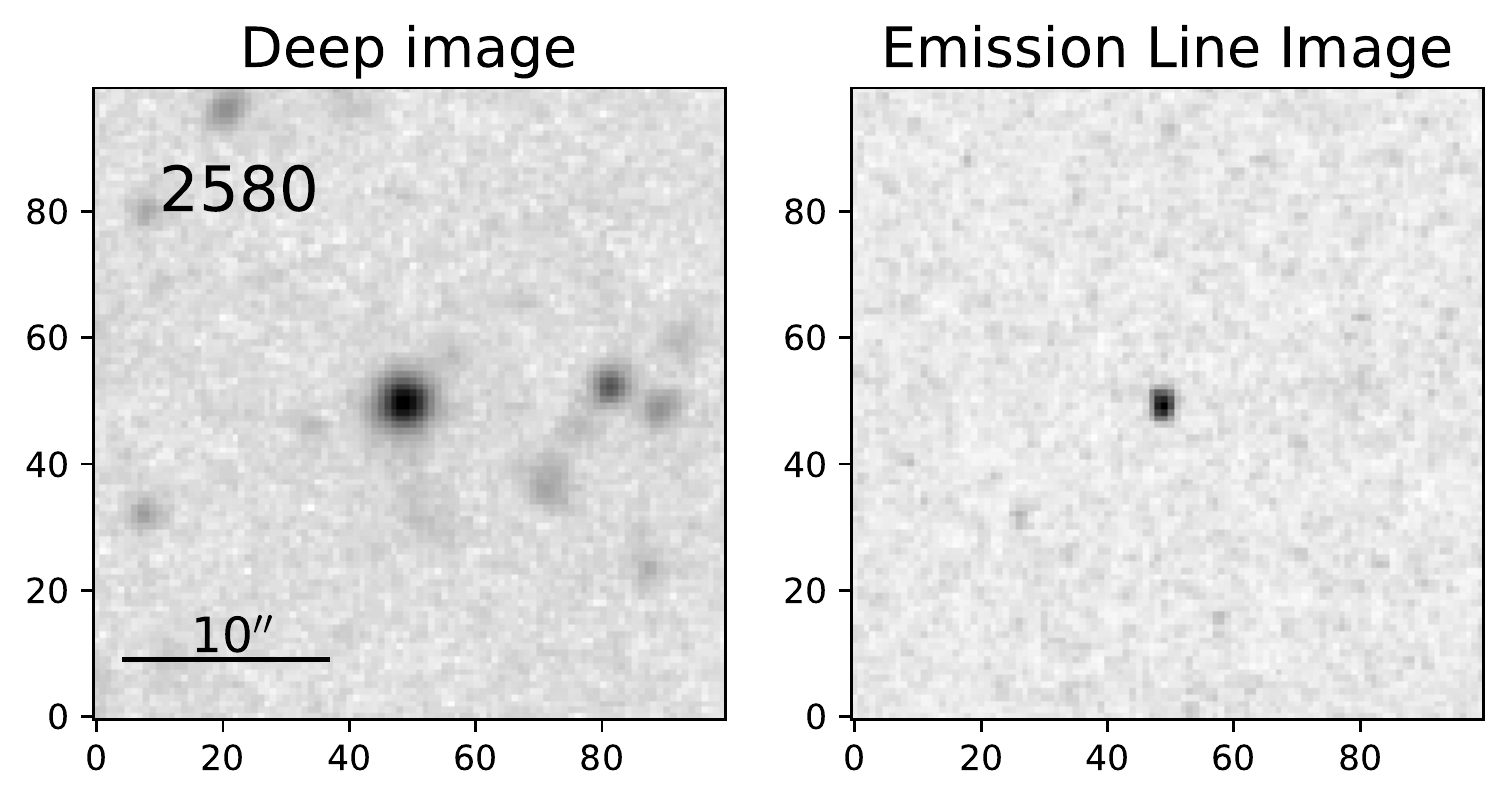}}\hfill
\subfigure{\includegraphics[width=3.6in]{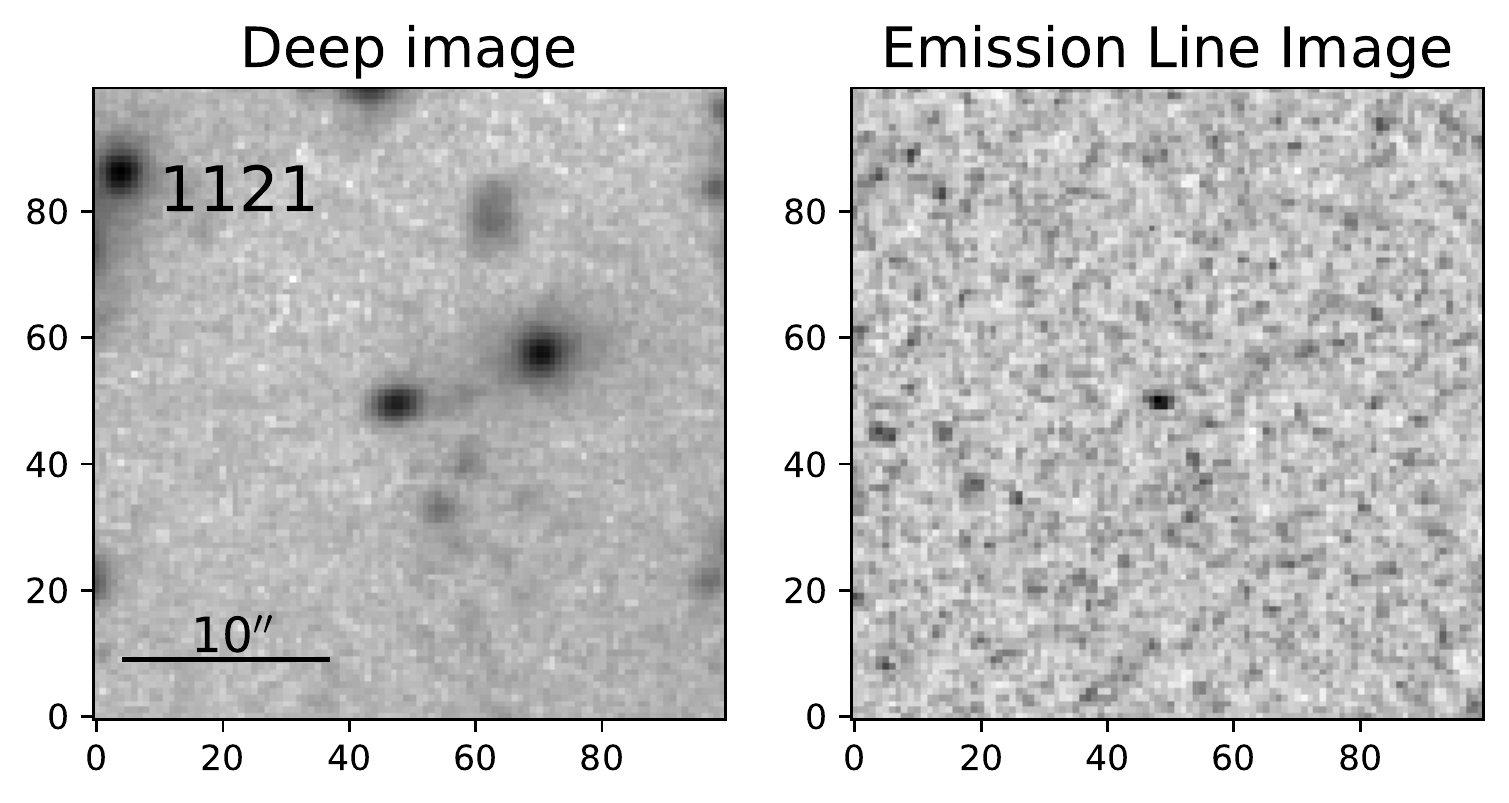}}}\caption{Cl0016+1609 postage stamp images. On the left, the deep image is shown. It is an addition of all the continuum light from the C2 SITELLE filter. On the right, a continuum-subtracted emission line image. 10$^{\prime\prime}= 64 \, $kpc. The first 6 of 13 sources are shown here.}\label{cl_asym1}
\end{figure*}

\begin{figure*}
\centering
\mbox{\subfigure{\includegraphics[width=3.6in]{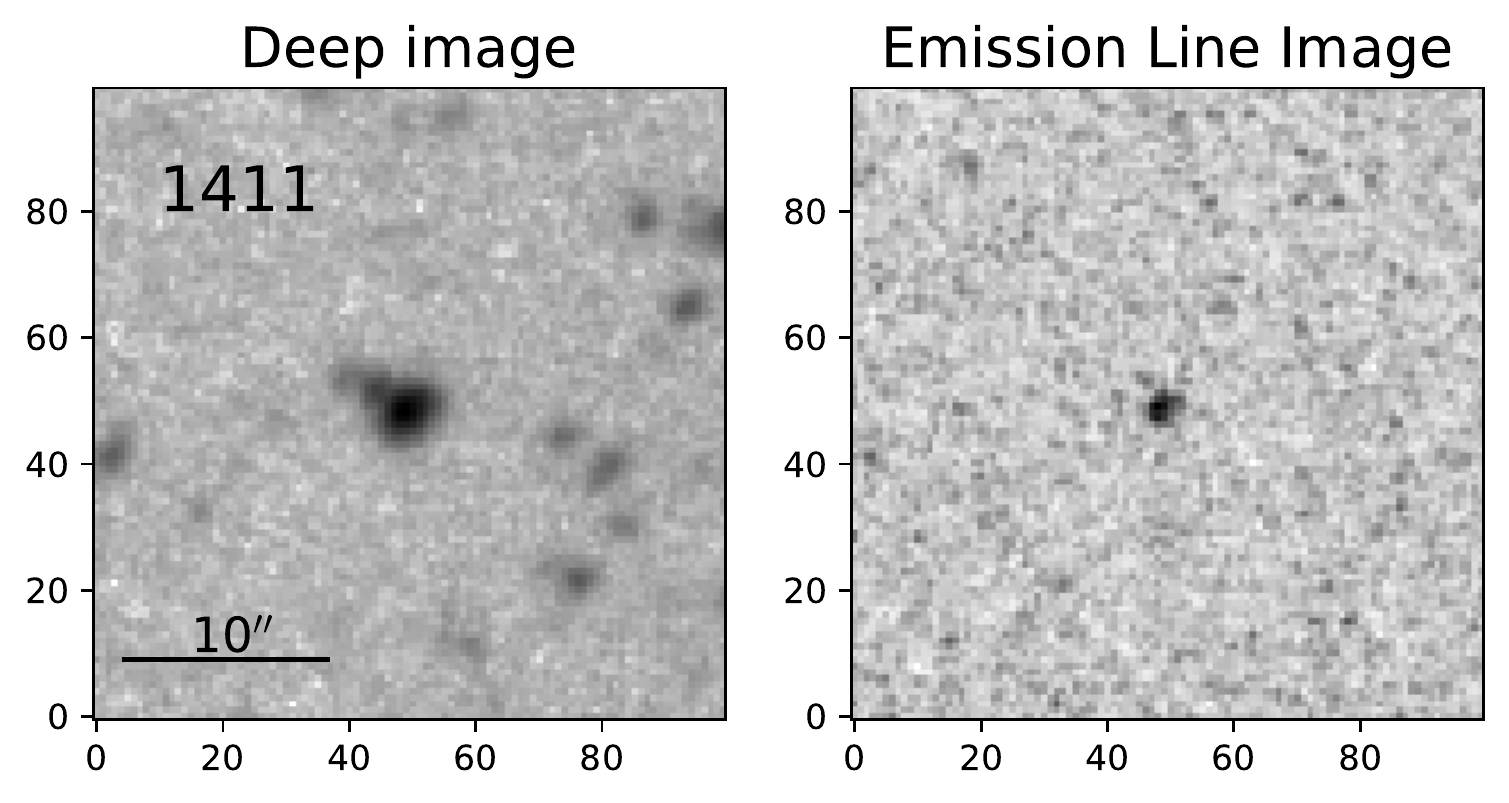}}\hfill
\subfigure{\includegraphics[width=3.6in]{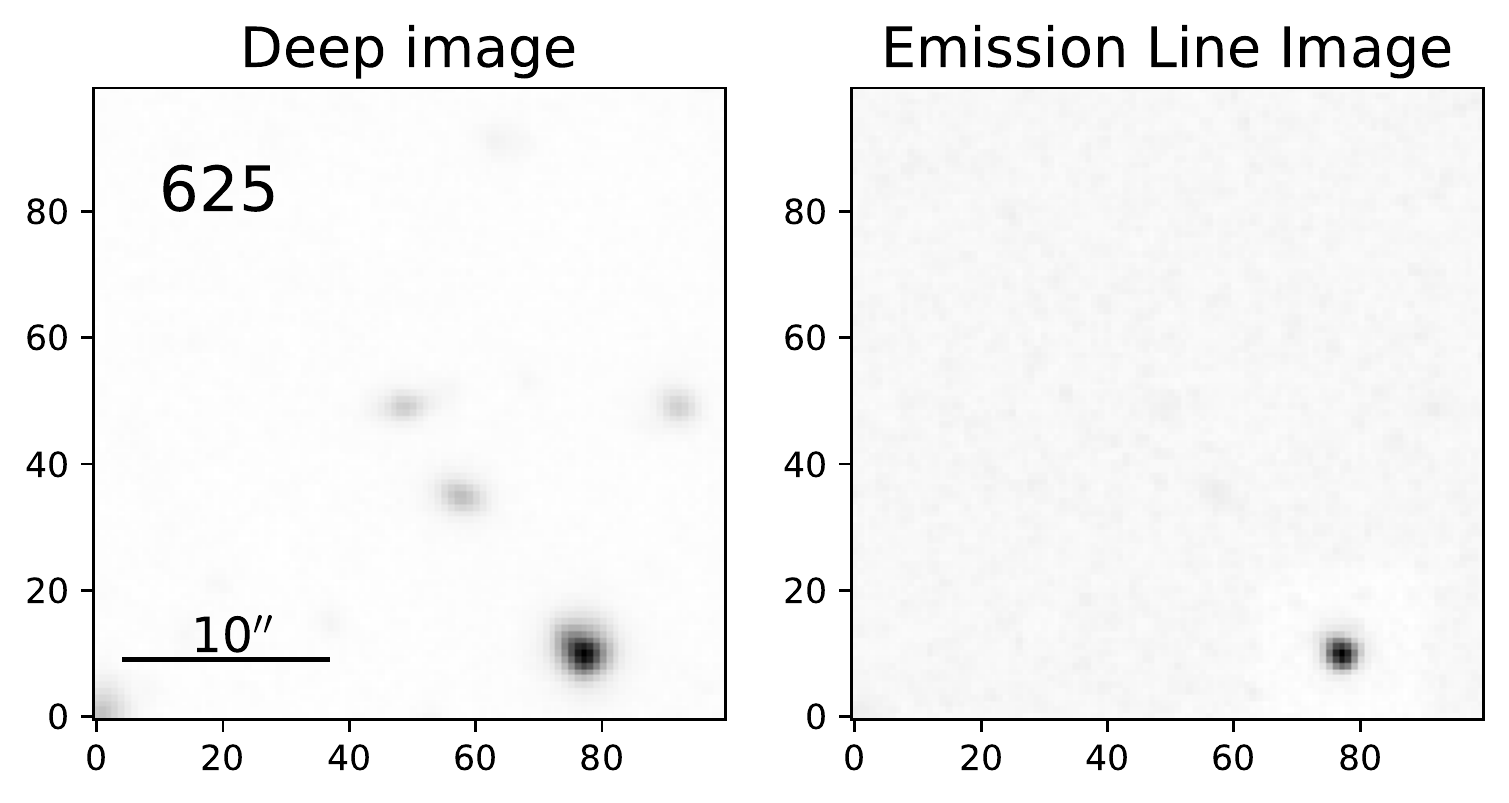}}}
\mbox{\subfigure{\includegraphics[width=3.6in]{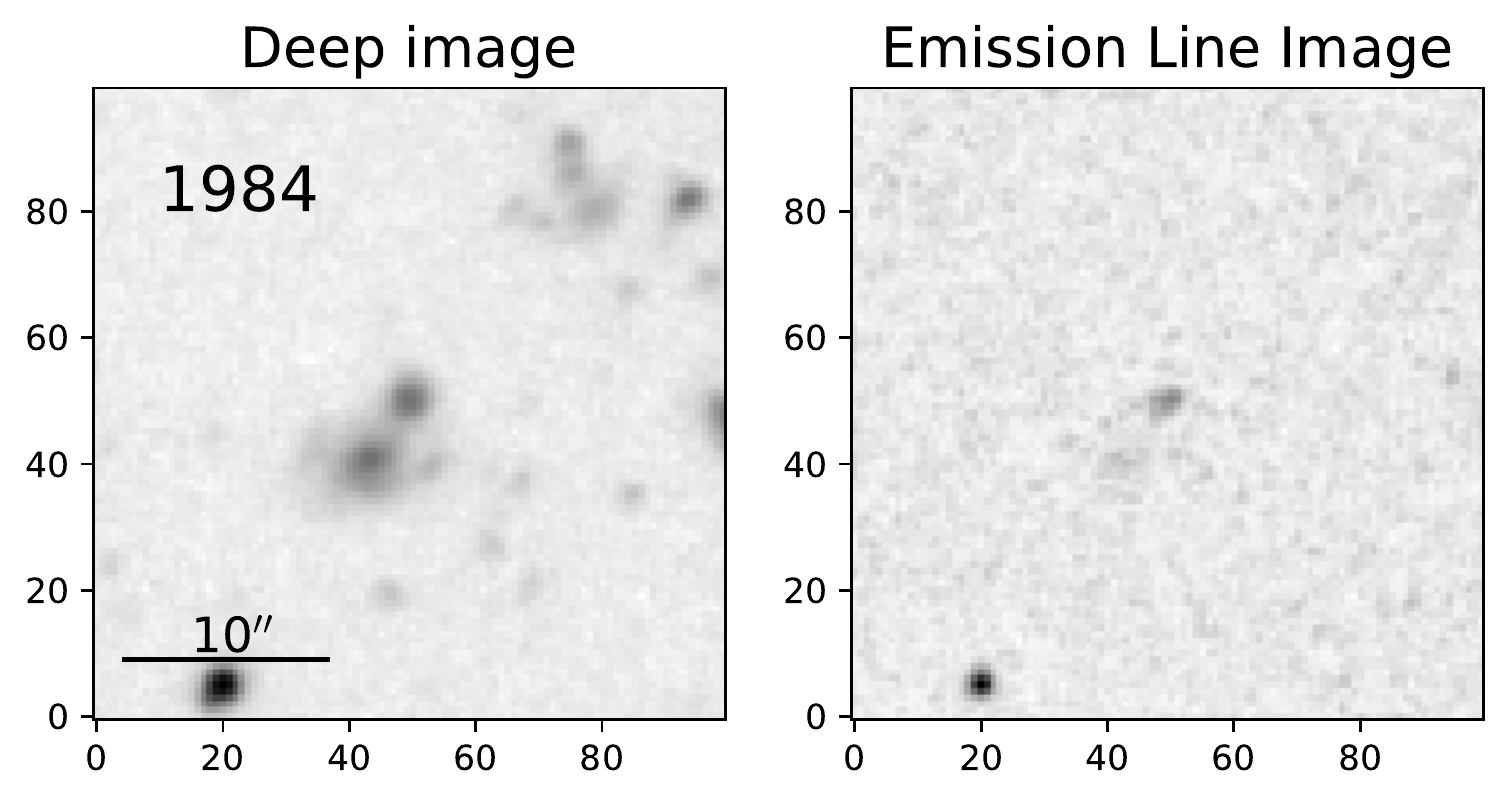}}\hfill
\subfigure{\includegraphics[width=3.6in]{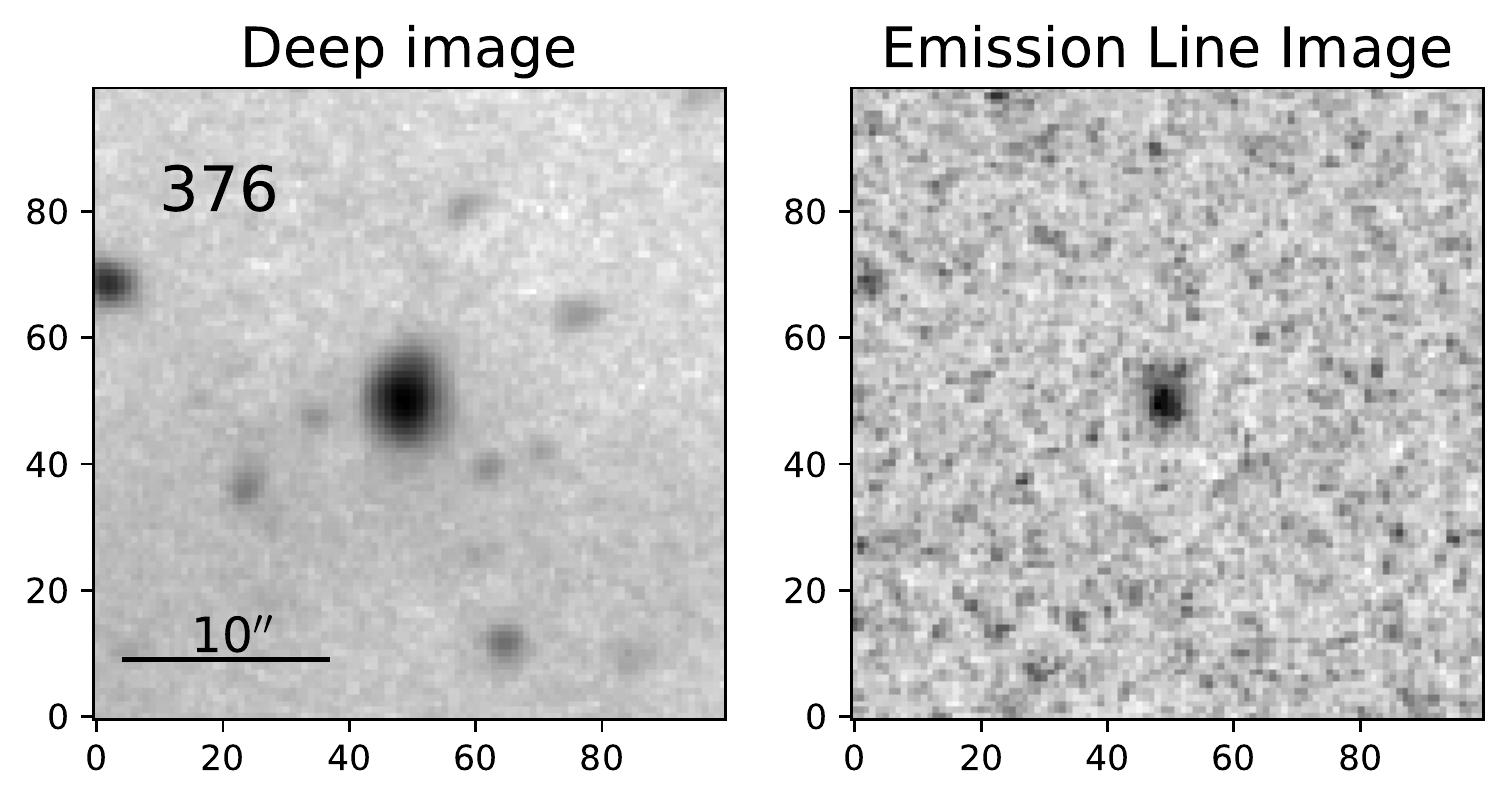}}}
\mbox{\subfigure{\includegraphics[width=3.6in]{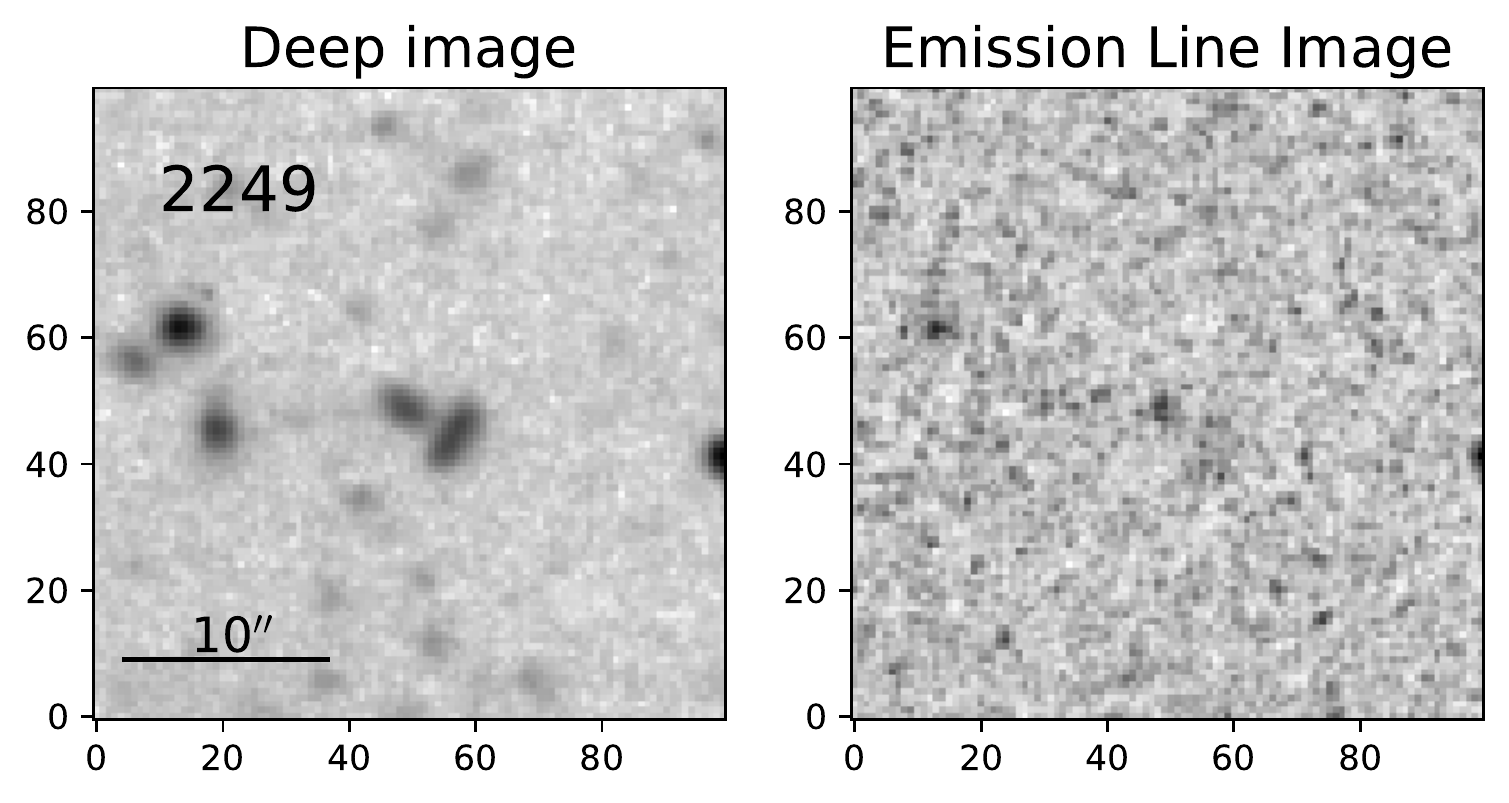}}\hfill
\subfigure{\includegraphics[width=3.6in]{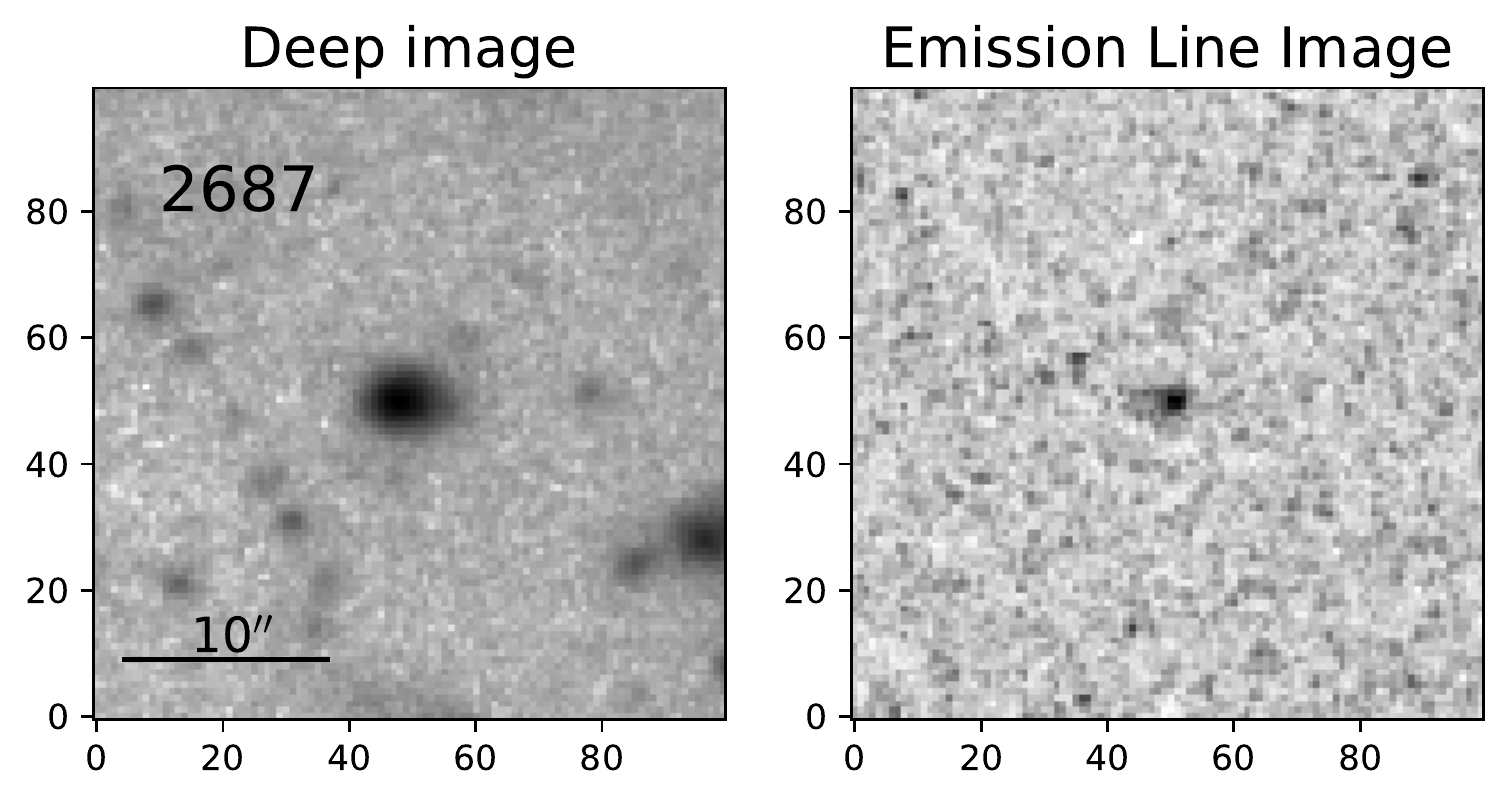}}}
\mbox{\subfigure{\includegraphics[width=3.6in]{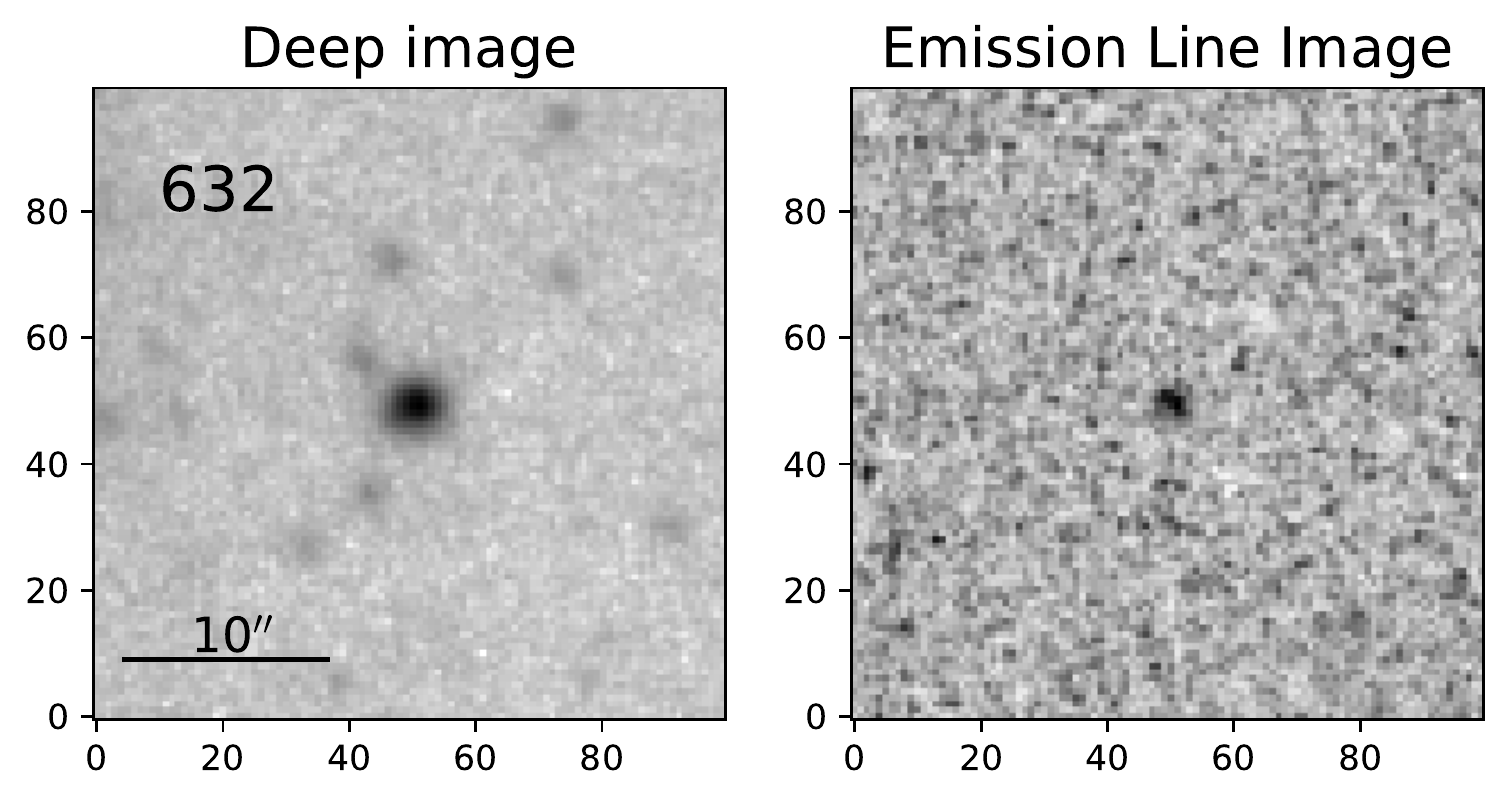}}\hfill}
\caption{Cl0016+1609 postage stamp images. On the left, the deep image is shown. It is an addition of all the continuum light from the C2 SITELLE filter. On the right, a continuum-subtracted emission line image. 10$^{\prime\prime}= 64 \, $kpc. The second 7 of 13 images are shown here.}\label{cl_asym2}
\end{figure*}

\begin{figure*}
\centering
\mbox{\subfigure{\includegraphics[width=3.6in]{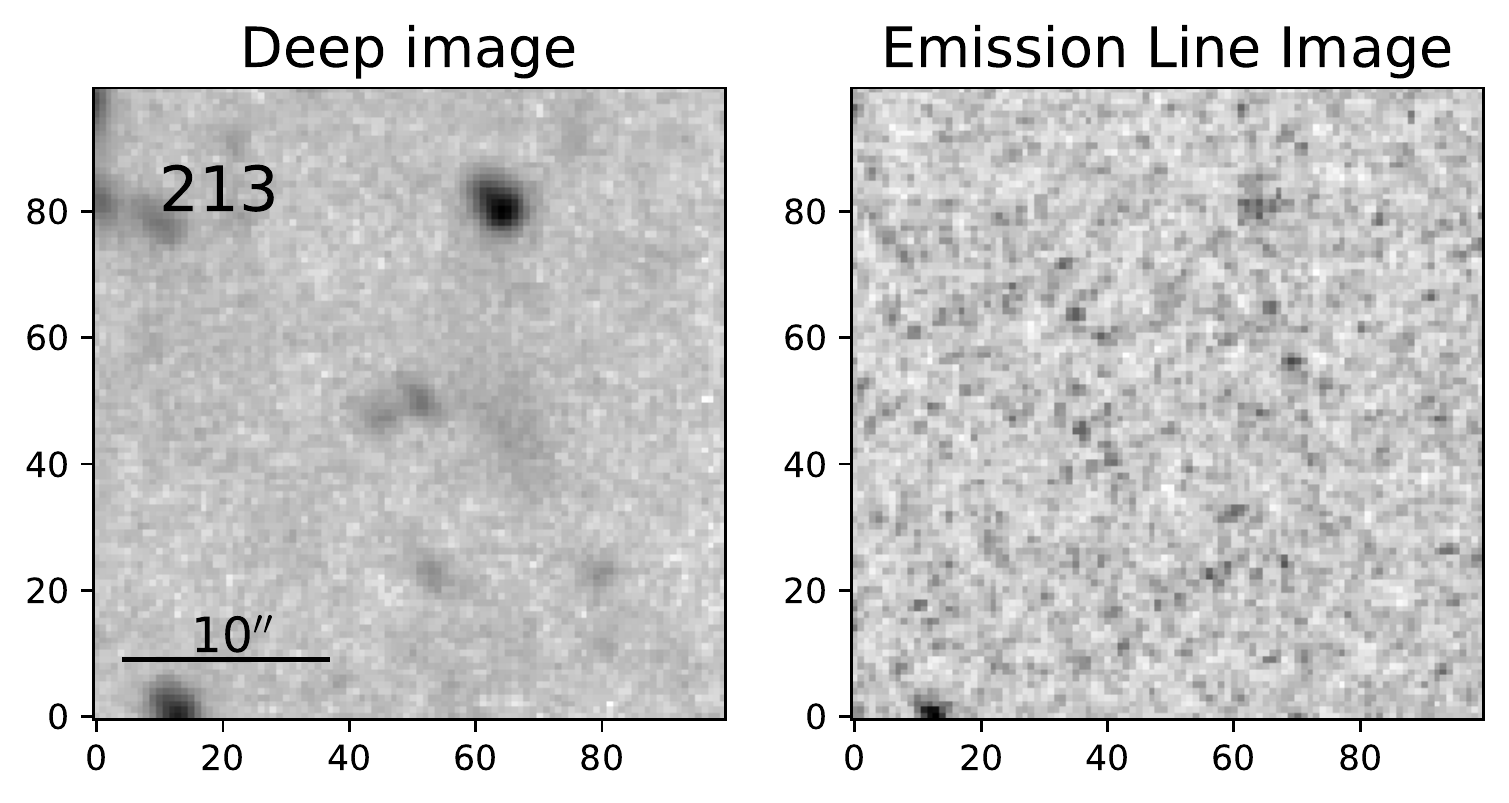}}\hfill
\subfigure{\includegraphics[width=3.6in]{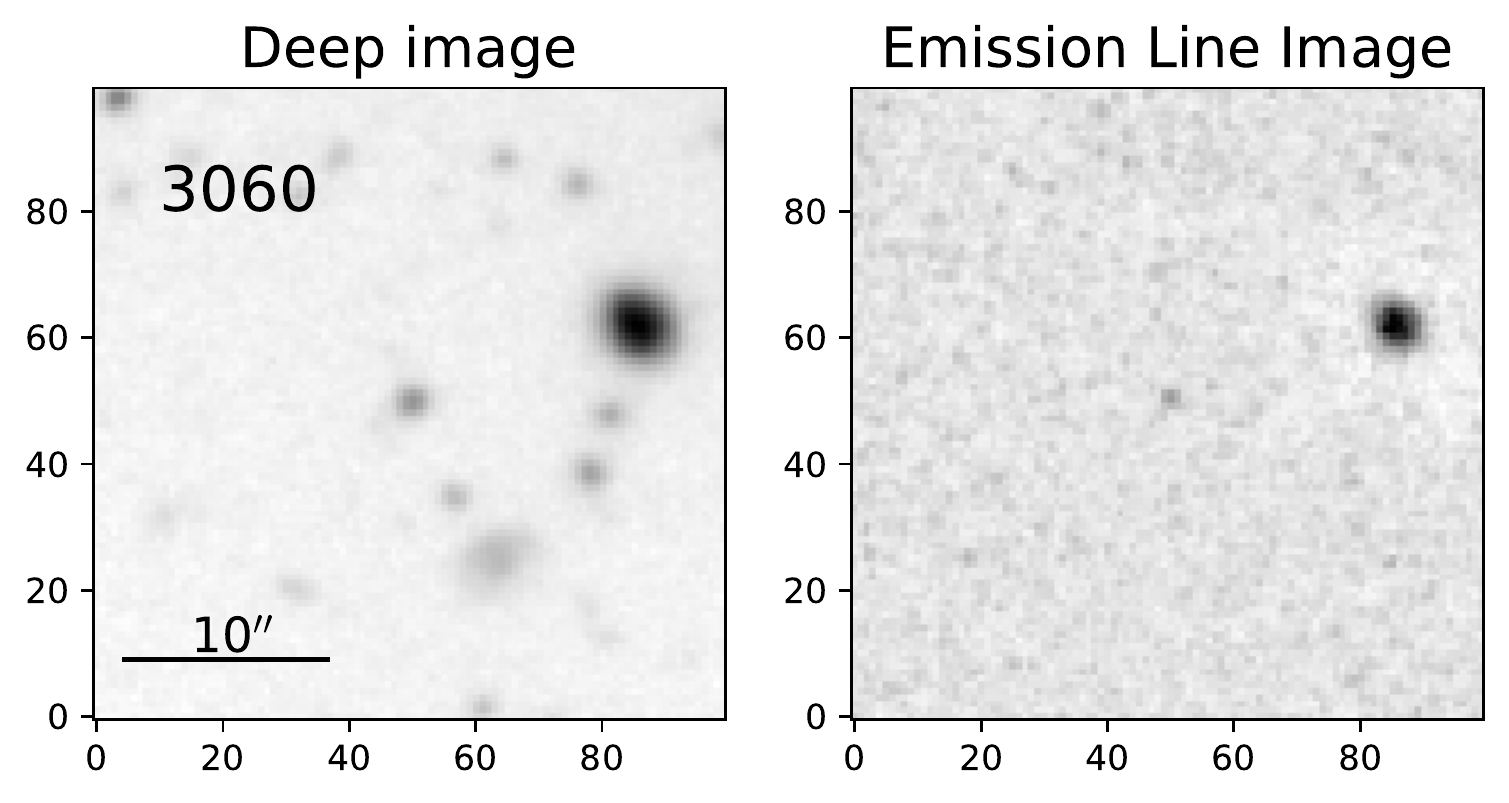}}}
\mbox{\subfigure{\includegraphics[width=3.6in]{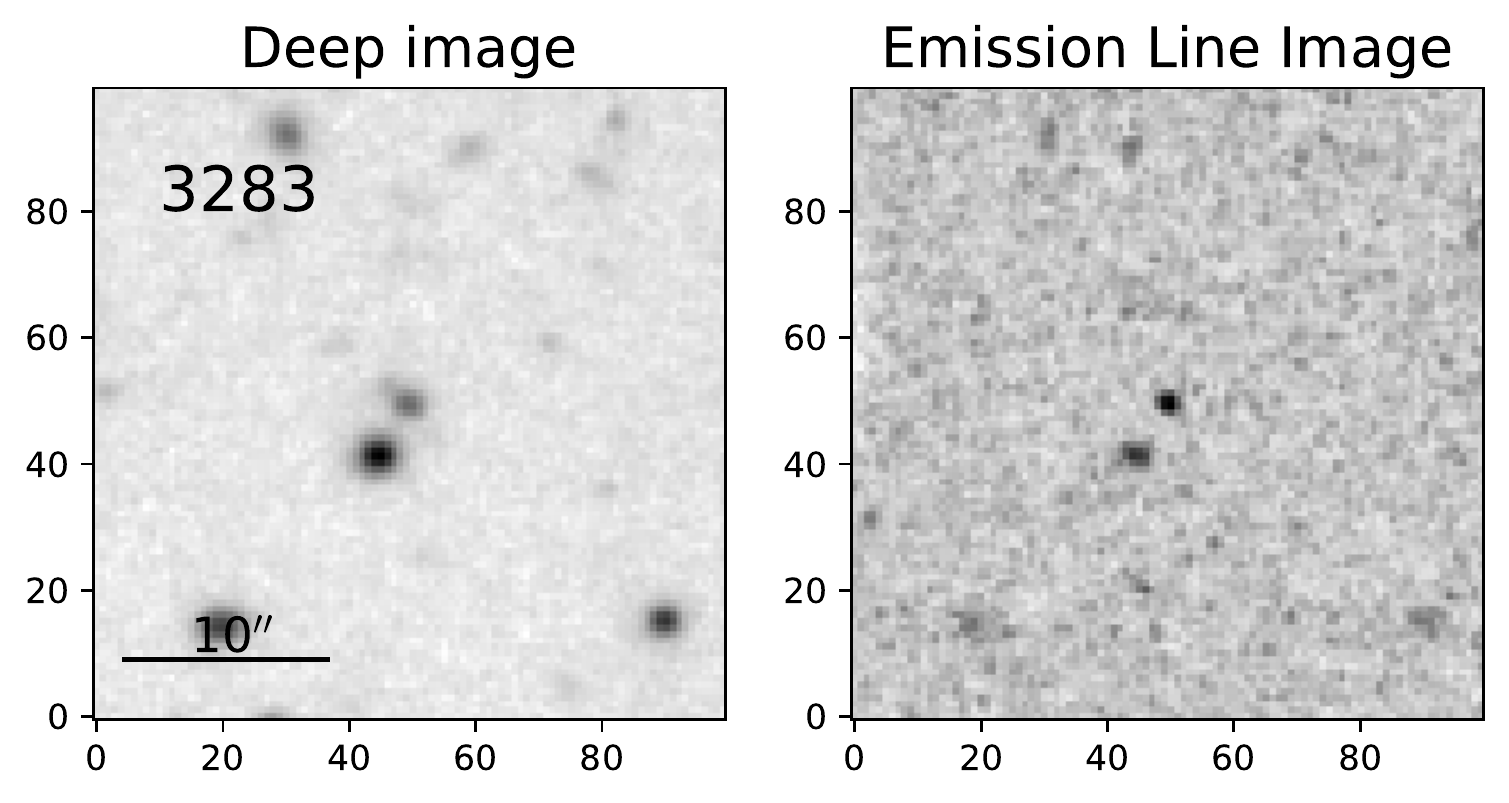}}\hfill
\subfigure{\includegraphics[width=3.6in]{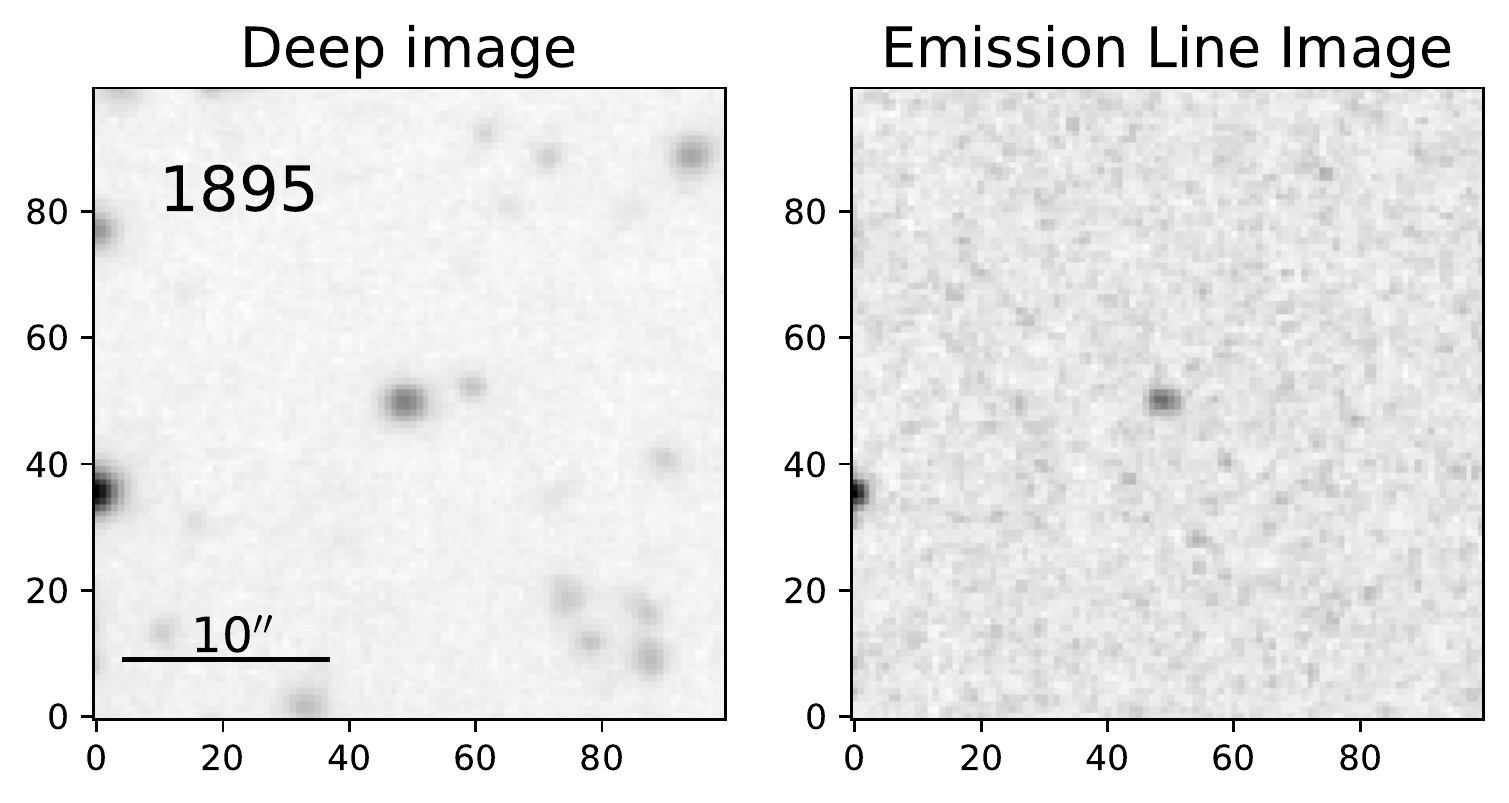}}}
\mbox{\subfigure{\includegraphics[width=3.6in]{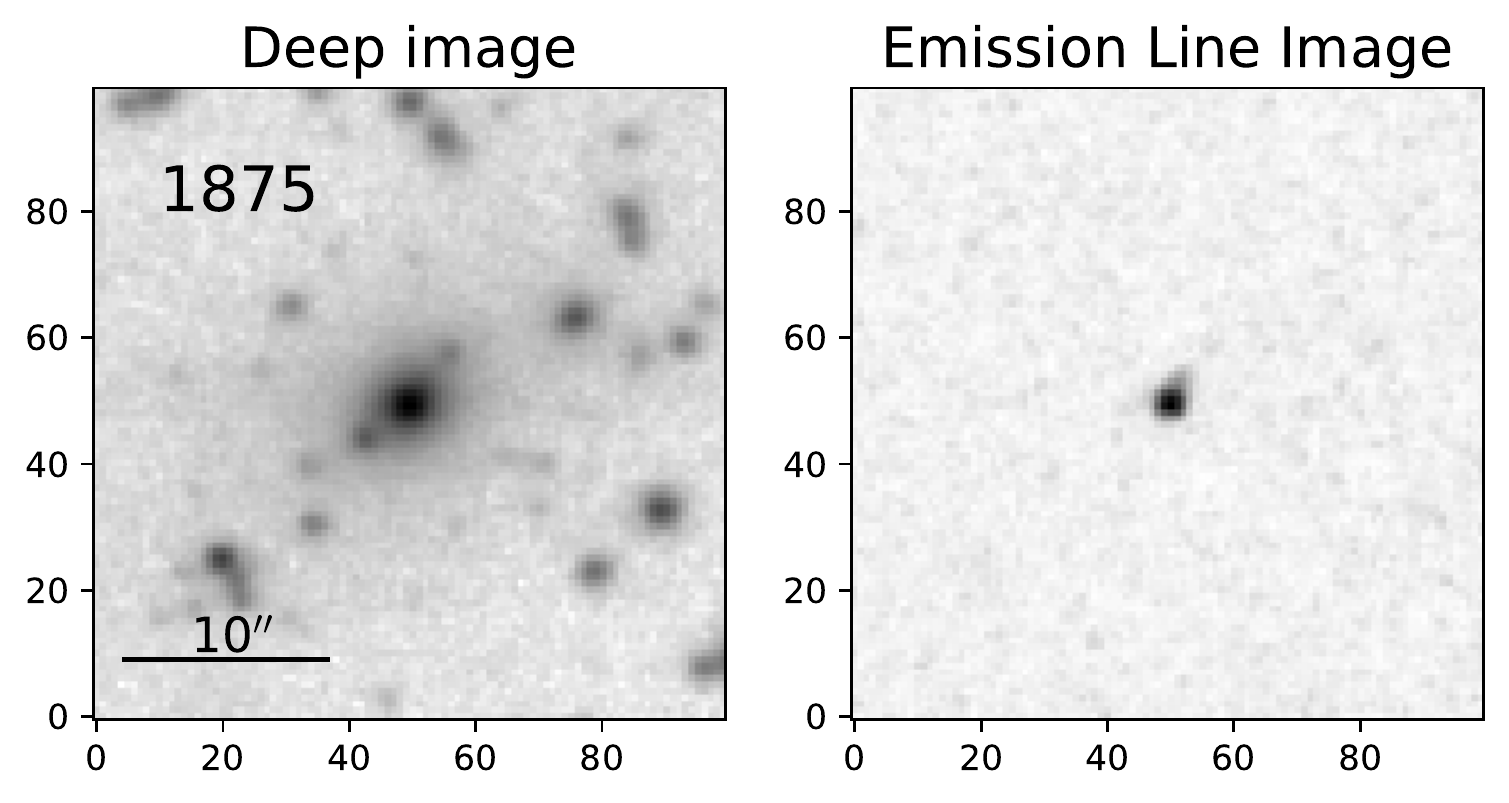}}\hfill
\subfigure{\includegraphics[width=3.6in]{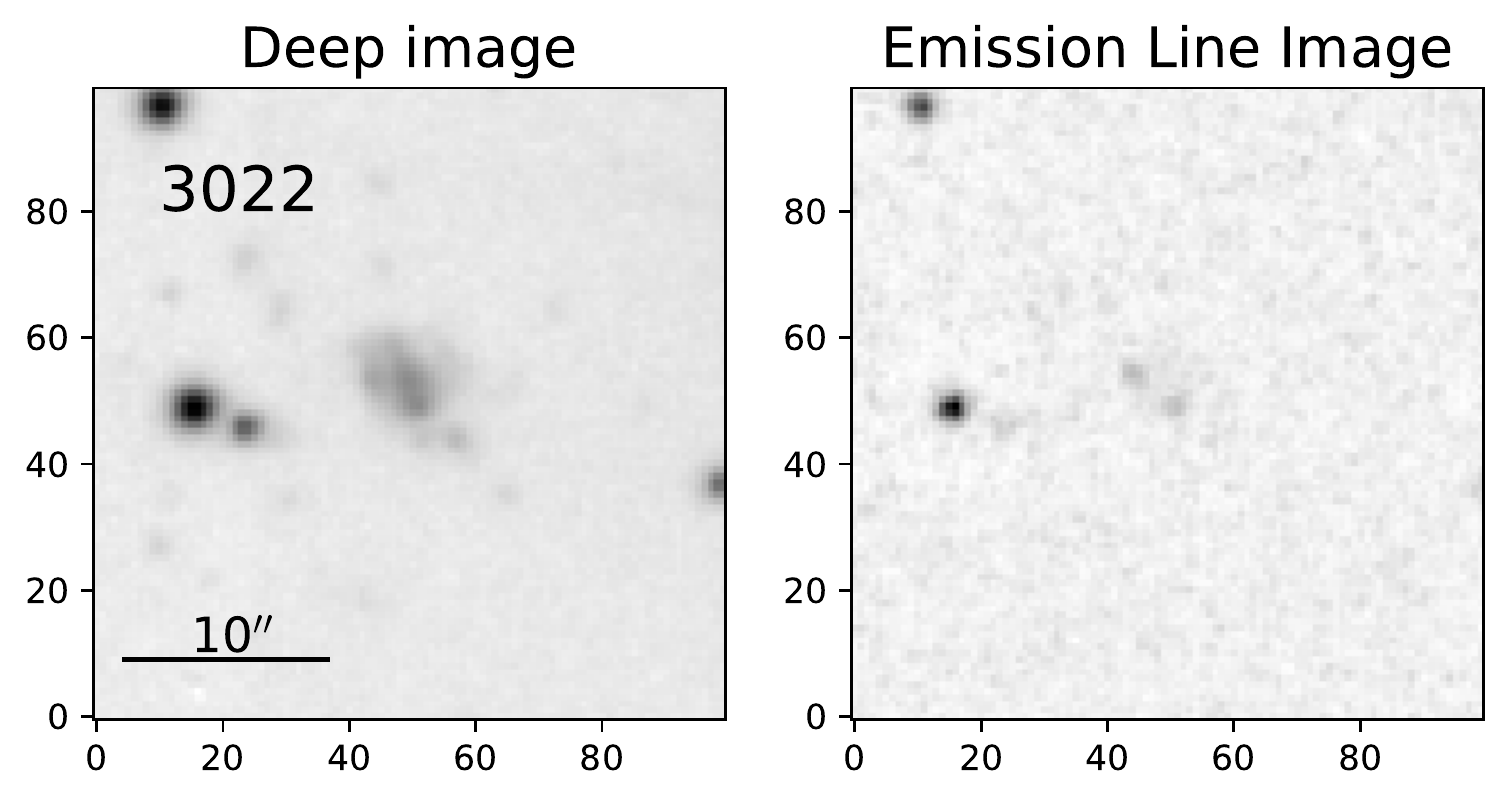}}}
\caption{MACSJ1621.4+3810 postage stamp images. On the left, the deep image is shown. It is an addition of all the continuum light from the C3 SITELLE filter. On the right, a continuum-subtracted emission line image. 10$^{\prime\prime}= 59 \, $kpc. The first 6 of 10 images are shown here.}\label{macs_asym1}
\end{figure*}

\begin{figure*}
\centering
\mbox{\subfigure{\includegraphics[width=3.6in]{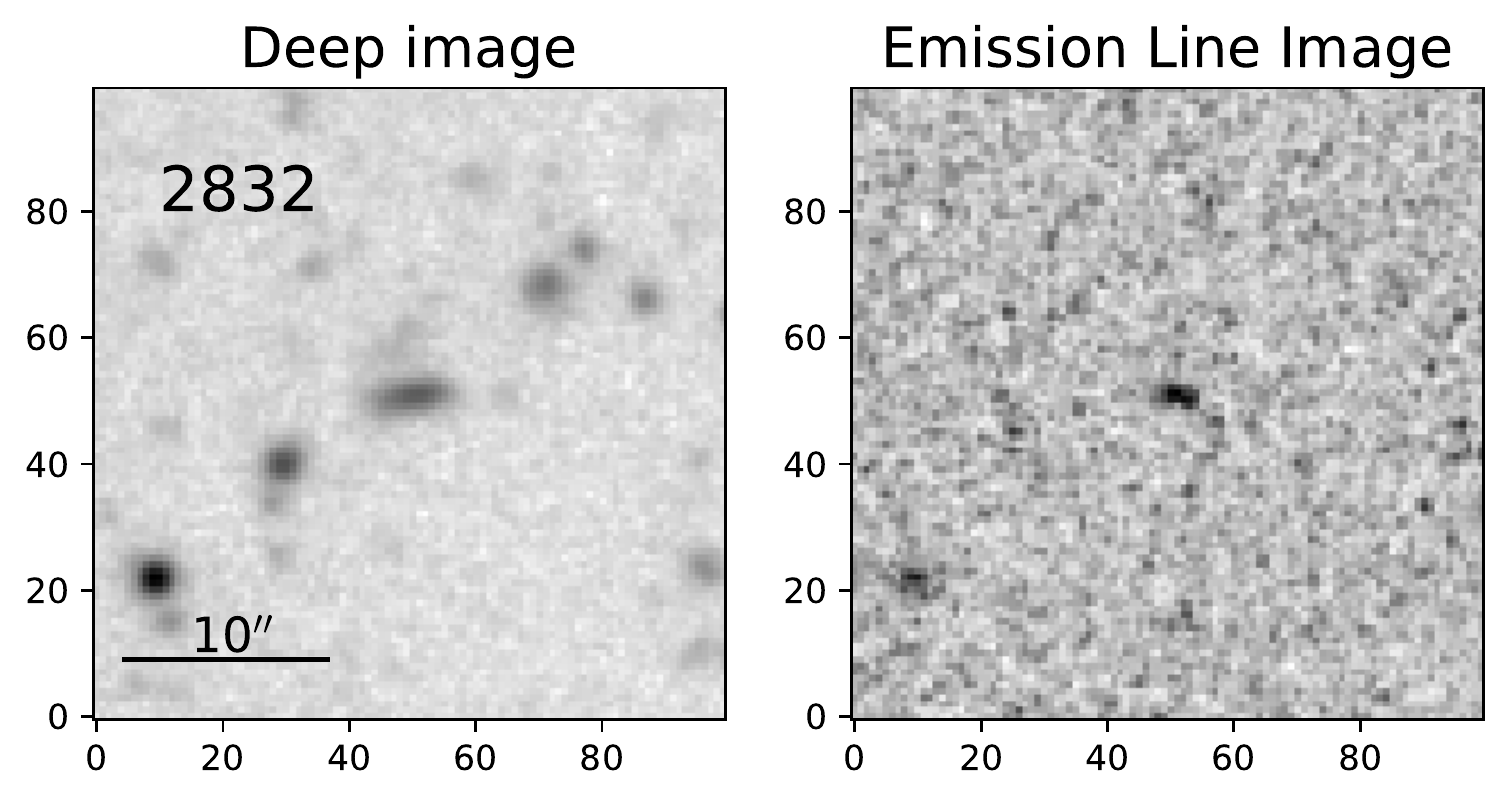}}\hfill
\subfigure{\includegraphics[width=3.6in]{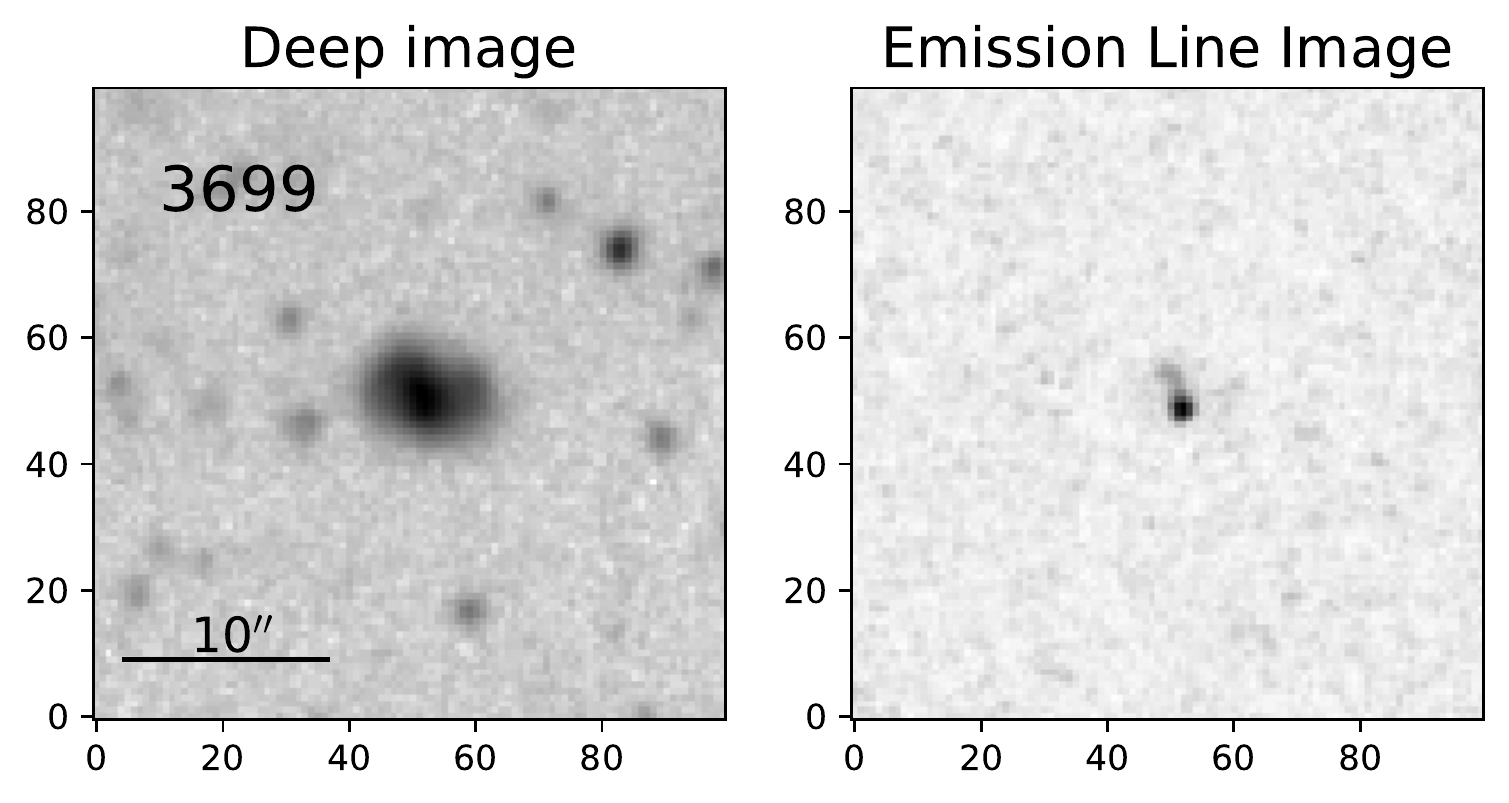}}}
\mbox{\subfigure{\includegraphics[width=3.6in]{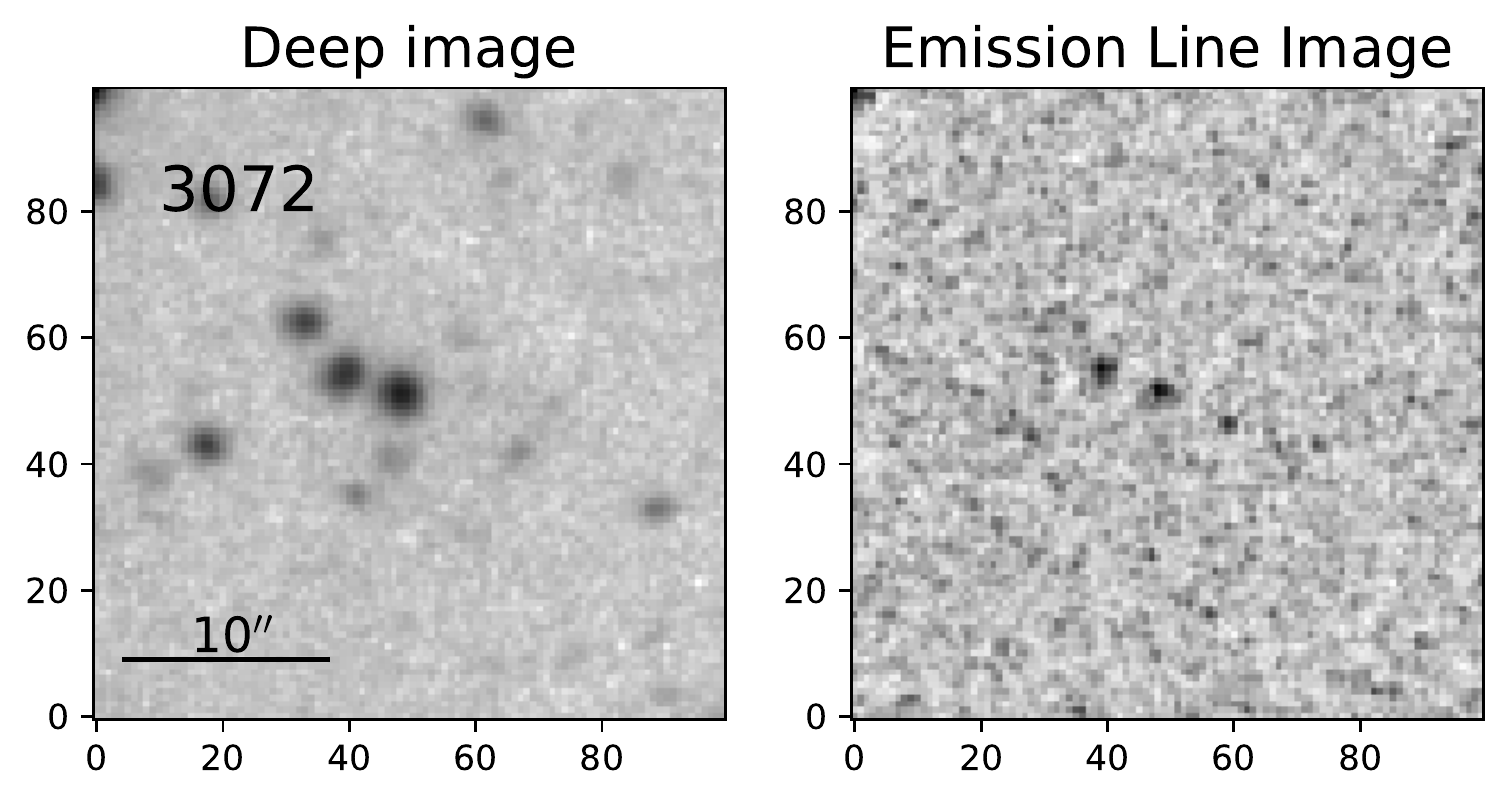}}\hfill
\subfigure{\includegraphics[width=3.6in]{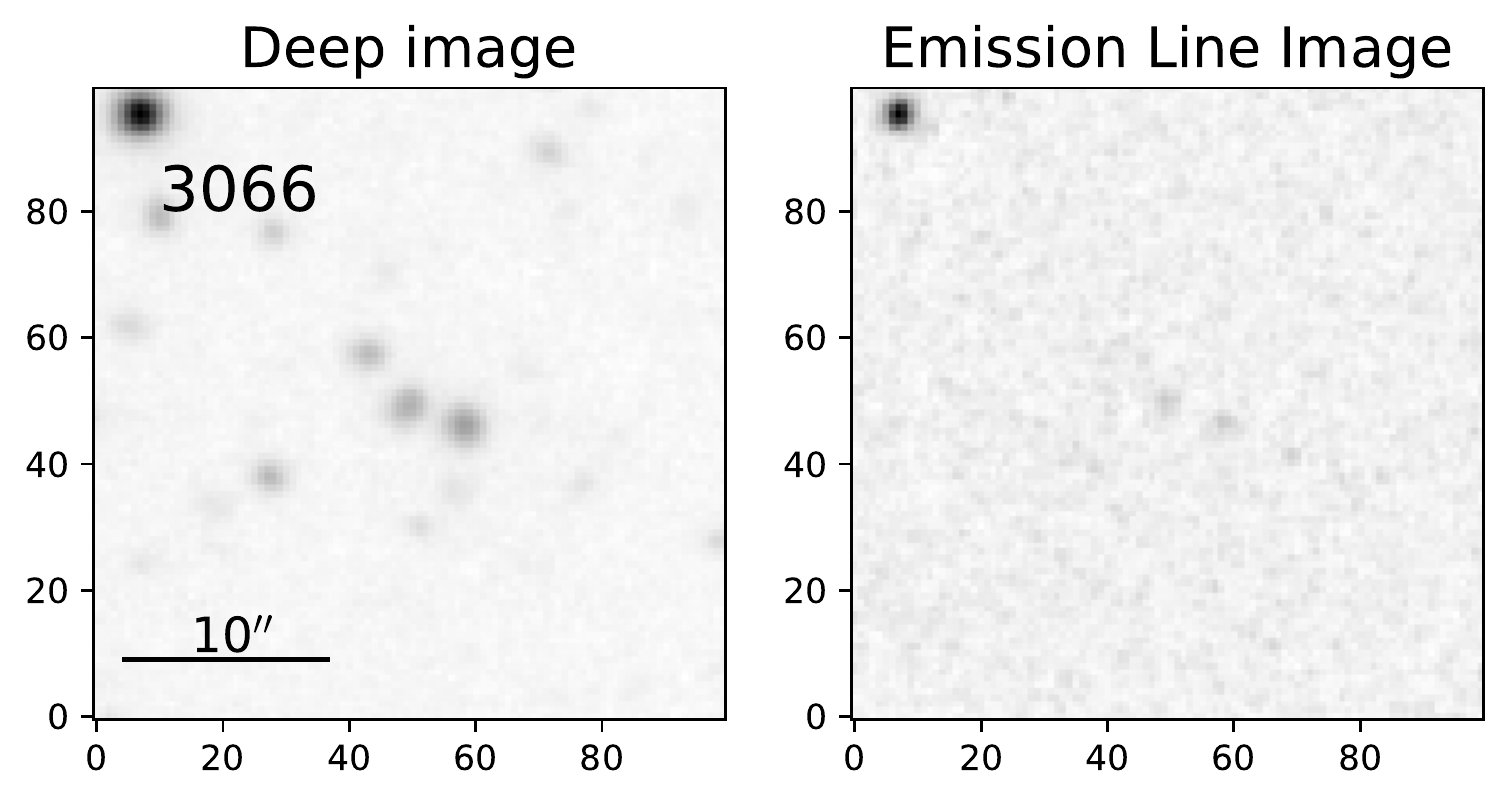}}}
\caption{MACSJ1621.4+3810 postage stamp images. On the left, the deep image is shown. It is an addition of all the continuum light from the C3 SITELLE filter. On the right, a continuum-subtracted emission line image. 10$^{\prime\prime}= 59 \, $kpc. The second 4 of 10 images are shown here.}\label{macs_asym2}
\end{figure*}

The Python program {\it statmorph} \citep{rod19} is used to further quantify the galaxy morphology, for all sources with emission line $S/N > 3$ and within the supercluster redshifts ($0.53-0.55$ for Cl0016+1609 and $0.46-0.47$ for MACSJ1621.4+3810). The {\it statmorph} code provides three shape paramenters explored in this study. They are, the Gini coefficient (G), the M$_{20}$ parameter (M$_{20}$) and the merger statistic. As discussed in \citet{lot04, lot08, sny15}, G is a measure of how the flux is distributed over the galaxy pixels. With G~$=1$, all the flux in concentrated in a single pixel, whereas with G~$=0$, the flux is equally distributed over all galaxy pixels. M$_{20}$ is the normalized second order moment of the brightest 20\% of the galaxy's flux. It traces bright features such as arms, bars, nuclei, bright star formation regions. The merger statistic separates merging galaxies in G - M$_{20}$ space, It is defined by the grey solid line in the left hand panels of Figure~\ref{cl_rprops}. Table~\ref{tab:objpropC} and Figure~\ref{cl_rprops} show the results for the star forming galaxies in the Cl0016+1609 and MACSJ1621.4+3810 superclusters.

For Cl0016+1609, the majority of emission line galaxies are  disky, with 69$^{+10}_{-15}$\%  as Sb/Sc/Sd/Irr systems, which can be seen in Figure~\ref{cl_rprops} (top), with one system (8$^{+14}_{-3}$\%) characterised as E/SO/Sa, and three (23$^{+15}_{-8}$\%) merging systems. 100$^{+0}_{-37}$\%  of the merging systems, shown as triangles, are found in moderate density environments (LGD$=3000-8000$), where the galaxy density is high enough that encounters would be expected.  ID 1411 is on the border of a positive merger statistic, it is identified as E/SO/Sa, thus it could be an emission line elliptical; however, with its blue color, S0/Sa is a more likely identification. No trend with orientation, position angle or galaxy size is found with either star formation rate, or with local galaxy density (not shown). 

For MACSJ1621.4+3810, the emission line galaxies are fairly evenly split between disky systems (circles, 40$^{+16}_{-12}$\%) and mergers (triangles, 60$^{+12}_{-16}$\%), with 0$^{+15}_{-0}$\% bulge-dominated systems, according to the G - M$_{20}$ classification scheme (see Figure~\ref{cl_rprops}, bottom). Even the BCG (ID 1875, with G~$=0.52$ and M$_{20}=-1.9$) is classified as disky, though it is not far from the E/S0/Sa border. And, recall, the deep SITELLE images are constructed from the fairly narrow C3 filter, which has a bandwidth of 500~\AA, which includes the flux of the line emission, and could be dominating the luminosity. 83$^{+7}_{-23}$\% of the merging systems are found in moderate density environments (between $\sim9000-25000$) and all have nearby (projected) companions as seen in the deep images of Figures~\ref{macs_asym1} and~\ref{macs_asym2}. Five of six (83$^{+7}_{-23}$\%) of the merging systems appear large, as their best-fit model includes the neighboring galaxy. The merging galaxies (ID: 213, 1895, 3066, 3283, 2832, 3072) have low to moderate star formation rates ($0.3-2.7\,$M$_\odot$yr$^{-1}$ ). No trend with orientation, or galaxy size is found for either the star formation rate or local density. The star forming mergers with merging characteristics are not separated in color, environment, size or magnitude from the non-merging star forming galaxies (not shown).  

\subsection{Size and shape of the line emission}

Several studies have shown that the disks of star forming galaxies become truncated as the galaxies move through the intracluster medium \citep[e.g.,][]{gul20}, caught by the cluster potential. Such environmentally driven gas depletion is expected to preferentially remove gas from outskirts \citep{gun72,bek09,vul20}. In this case, one would expect to see the galaxies in more dense environments have a smaller extent of emission line gas relative to their host galaxy's size. To investigate, the Petrosian radius for both the deep galaxy image ($R_{gal}$) and emission line image ($R_{em}$) from SITELLE are measured using SExtractor. The ratio of the two is shown in the middle panels of Figure~\ref{cl_rprops}. Values of $R_{gal}/R_{em}$ are also listed in Table~\ref{tab:objpropC} and range from $\sim 0.5 ~- \sim 1.5$. These values are similar to the range found in \citet{vul19}, who examine four star forming galaxies inside of local filaments \citep{Tempel14}. They find ratios between $0.5-2$, where the emission line image is of H$\alpha$ emission from MUSE observations.  No strong trend with local galaxy density is found for Cl0016+1609, but a least squares fit to the data for MACSJ1621.4+3810 finds a trend with R$^{2}$ statistic of 0.65. 

\section{Frequency of emission line galaxies} 

SITELLE is a very efficient instrument for finding emission line galaxies in wide fields. The deep $i^{\prime}$ MegaCam images from CFHT for Cl0016+1609 list 135568 objects, 24747 with magnitudes brighter than 22.5. There are simply too many to observe individually. The red sequence method is a great way to identify the 5816 cluster members with $i^{\prime}<25$, but misses the blue - emission line - galaxies identified in the present study. Even attempting to use $g^{\prime}-i^{\prime}$ color to take spectra of the blue galaxies, would require observing 42821 galaxies. If an observer wanted to find these galaxies, but did not have access to SITELLE, they would have to observe the 129 bright blue galaxies spectroscopically. A program using a multi-object spectrograph would work, but would be quite inefficient, finding only $\sim$10\% of the galaxies had strong emission lines.

For MACSJ1621.4+3810, 10 emission line galaxies are identified, compared to the 3648 red sequence galaxies. 


\section{Conclusions}

For the z$\sim 0.5$ galaxies studied here, 79$^{+5}_{-9}$\% are found in moderate density regions (the main filaments), avoiding the main cluster and subcluster cores. This is what has been found at $z<0.3$ \citep{Edwards10, gal09,Bianconi16}. The  median  star formation rate for Cl0016+1609 galaxies is $\sim 2\,$M$_{\odot}$~yr$^{-1}$, and that for the slightly closer, less massive cluster MACSJ1621.4+3810, is $\sim 1\,$M$_{\odot}$~yr$^{-1}$. These values compare well to the $z\sim0.5$ value of the  [OII]~3727\AA~- derived rate of $1.6\,$M$_{\odot}$~yr$^{-1}$, calculated for $h=0.7$ and measured by \citet{coo08}. The emission line galaxies are blue and have disky/Irregular shapes.

The most obvious trend found in our data is with galaxy stellar mass, with higher specific star formation rates found for the dwarf galaxies. Previous studies with the GAMA survey \citep{Alpaslan16} have found that the large-scale environment can modulate star formation rates. 
If the increased frequency of star formating galaxies in the moderate density environments is indeed a result of galaxy-galaxy interactions, these galaxies might be expected to show tell-tail  signs of merging like tidal tails or trainwreck shapes. Deep images from the SITELLE data of all the emission line sources are constructed and compared to images from within the line emission only. This reveals several cases of close companions, multiple cores and material connecting to nearby systems. A significant fraction (39$^{+9}_{-8}$\%) of emission line galaxies are characterized as merging systems, based on the G - M$_{20}$ merger statistic.

Using SITELLE is an efficient way to find emission line galaxies, even at redshifts of 0.5. These galaxies trace the a priori identified supercluster filaments. In the near future, when photometric filament finding algorithms produce large databases of potential cluster scale filaments, SITELLE would be an ideal instrument to use to confirm and study the candidates.

\acknowledgments

L. O. V. E. acknowledges the California Polytechnic State University, San Luis Obispo, Physics department for start up funds which supported travel to IAP, and undergraduate research support programs for K. Zhang and J. Fraga. K.A.E. bravely oversaw W.E. and newly born Skyler Beasley-Murray during the IAP trip. F. D. acknowledges long-term support from CNES. I. M. acknowledges financial support from the State Agency for Research of the Spanish MCIU through the ``Center of Excellence Severo Ochoa" award to the Instituto de Astrof\'\i sica de 
Andaluc\'\i a (SEV-2017-0709) and the project PID2019-106027GB-C41.

\vspace{5mm}
\facilities{CFHT(SITELLE)} Based on observations obtained with SITELLE, a joint project between Université Laval, ABB-Bomem, Université de Montréal and the CFHT with funding support from the Canada Foundation for Innovation (CFI), the National Sciences and Engineering Research Council of Canada (NSERC), Fond de Recheche du Québec - Nature et Technologies (FRQNT) and CFHT. 
\facilities{NED} This research has made use of the NASA/IPAC Extragalactic Database (NED),
which is operated by the Jet Propulsion Laboratory, California Institute of Technology,
under contract with the National Aeronautics and Space Administration.
\facilities{CFHT(MegaPrime/MegaCam)}Based on observations obtained with MegaPrime/MegaCam, a joint project of CFHT and CEA/IRFU, at the Canada-France-Hawaii Telescope (CFHT) which is operated by the National Research Council (NRC) of Canada, the Institut National des Science de l'Univers of the Centre National de la Recherche Scientifique (CNRS) of France, and the University of Hawaii. This work is based in part on data products produced at Terapix available at the Canadian Astronomy Data Centre as part of the Canada-France-Hawaii Telescope Legacy Survey, a collaborative project of NRC and CNRS.
\facilities{Subaru(HSC)}Based in part on data collected at Subaru Telescope, which is operated by the National Astronomical Observatory of Japan.

\software{astropy \citep{2013A&A...558A..33A},  
          SExtractor \citep{Bertin96}
          } 
\newpage

\bibliography{sample63}{}
\bibliographystyle{aasjournal}

\end{document}